\documentclass[ aip, amsmath,amssymb, reprint,longbibliography]{revtex4-1}
\usepackage{comment}
\usepackage{physics}
\usepackage{float}
\usepackage{amsmath}
\usepackage{mathptmx,bm}
\usepackage{graphicx}
\usepackage{dcolumn}
\usepackage{tabularx}
\usepackage{epsf}
\usepackage{color}
\usepackage{hyperref}
\hypersetup{breaklinks,colorlinks,linkcolor=blue,citecolor=blue,urlcolor=blue}

\usepackage{epstopdf}%
\usepackage{verbatim}
\usepackage[english]{babel}
\newcommand{\bea}{\begin{eqnarray}}
\newcommand{\eea}{\end{eqnarray}}
\usepackage{amsfonts}
\usepackage{tikz}

\usepackage{pgfplots}

\DeclareMathAlphabet{\mathcal}{OMS}{cmsy}{m}{n}

\begin{document}

%\title{V-model in a nonequilibrium steady-state: quantum coherences, populations, and heat transport}intereference

%\title{Quantum coherences in thermal energy transport: The V-model as a case study}
\title{Quantum coherence-control of thermal energy transport: The V model as a case study}
%\title{Quantum control of thermal energy transport through intereference effects}
%transient and steady state behavior}
%OR
%\title{Quantum coherence and thermal energy transport: The V-model as a case study}

\author{Felix Ivander}
\affiliation{Chemical Physics Theory Group, Department of Chemistry and Centre for Quantum Information and Quantum Control,
University of Toronto, 80 Saint George St., Toronto, Ontario M5S 3H6, Canada}

\author{Nicholas Anto-Sztrikacs}
\affiliation{Department of Physics, University of Toronto, 60 Saint George St., Toronto, Ontario M5S 1A7, Canada}

\author{Dvira Segal}
\affiliation{Chemical Physics Theory Group, Department of Chemistry and Centre for Quantum Information and Quantum Control,
University of Toronto, 80 Saint George St., Toronto, Ontario M5S 3H6, Canada}
\affiliation{Department of Physics, University of Toronto, 60 Saint George St., Toronto, Ontario M5S 1A7, Canada}
\email{dvira.segal@utoronto.ca}

\date{\today}

\begin{abstract}
Whether genuine quantum effects, particularly quantum coherences, can offer an advantage to quantum devices is a topic of much interest.
Here, we study a minimal model, the three-level V system coupled to two heat baths, and investigate the role of quantum coherences in heat transport in both the transient regime and in the nonequilibrium steady-state. In our model, energy is exchanged between the baths through two parallel pathways, which can be made distinct through the nondegeneracy of excited levels (energy splitting $\Delta$) and a control parameter $\alpha$, which adjusts the strength of one of the arms. 
Using a nonsecular quantum master equation of Redfield form, we succeed in deriving closed-form expressions for 
the quantum coherences and the heat current in the steady state limit for closely degenerate excited levels. By including three ingredients in our analysis:
nonequilibrium baths, nondegeneracy of levels, and asymmetry of pathways,
we show that quantum coherences are generated and sustained  in the V model in the steady-state limit if three conditions, conjoining thermal and coherent effects are {\it simultaneously} met:
(i) The two baths are held at different temperatures. 
(ii) Bath-induced pathways do not interfere destructively.
(iii) Thermal rates do not mingle with the control parameter $\alpha$ to destroy interferences through an effective local equilibrium condition.
%DD3 note rephrasing "local equilibrium condition' rather than equilibrium state, since there is current when kh=alpha k_c.
Particularly, we find that coherences are maximized when the heat current is suppressed.
%When these conditions are simultaneously satisfied, quantum coherences and heat currents are completely out-of-phase. 
%We find the non-Secular Redfield Quantum Master equation approach used in this work faithfully captures regimes where population inversion and coherent-population-trapping (CPT) occur. 
On the other hand, the secular Redfield quantum master equation is shown to fail in a broad range of parameters. 
%where destructive interference takes place. % DDD2
Although  we mainly focus on analytical results in the steady state limit, 
numerical simulations reveal that the transient behavior of coherences {\it contrasts} the steady-state limit:
Large long-lived transient coherences vanish at steady state, while weak short-lived transient coherences survive, suggesting that different mechanisms are at play in these two regimes.
%Indeed,  while the duration of the transient dynamics is critically determined by the excited state energy splitting, $\Delta$, the steady state values of coherences, populations and the current only weakly depend on this parameter.
Enhancing either the lifetime of  transient coherences or their magnitude at steady state thus requires the control and optimization of different physical parameters. 
%a trade-off relationship. 
%Nontrivial effects emerge in transitive regions and transient quantum coherences are in the contrary benefiting heat transport properties.
\end{abstract}

\maketitle

\section{Introduction}

%Intro: Quantum coherences and steady-state heat transport
Quantum coherences are central to many fields of research, most pertinently to quantum information processing where  maintaining quantum coherences is crucial for operation \cite{chuang}. 
% DDD what types of quesions in biology? light harvesting?
In biology, numerous studies addressed the question of whether quantum coherences are a resource in natural processes %cite{ph1,QB,P2,Plenioreview,AC1,AC2,P1,P2,Engel2007,O1,O2,O3,V1,V2,bio1}.
such as photosynthesis\cite{Engel2007,Vaziri_2010,ph1,Plenioreview,JYZJPCL,QB,Dwayne17}, 
 %%%FFF removed citation P2
 avian navigation \cite{AC2,AC1,AC3},
 %FFF AC1 focuses more on entanglement but still ok
 %protein structure\cite{P1,P2,Engel2007,Plenioreview}, 
 %olfaction \cite{O1,O2,O3},
 %FFF removed olfaction: quantum effects but inelastic tunneling/spectroscopy not coherence
 and vision \cite{Dwayne06,V2,V1}.
 %%% FFF new effect but dodgy
 %and consciousness \cite{con1,con2}
In quantum thermal machines, quantum coherences can lead to, e.g., 
% \cite{Latune2021},  % DDD?
negative entropy \cite{NE}, violation of detailed balance \cite{db1,db2}, work extraction from a single bath \cite{SW3,SW1,SW2}, 
enhancement of fluctuations \cite{jliu2021},
and boosts to performance through various means \cite{db1,DBc2,QTMC2,COHQAR2,JCao2,COHQAR1,COHQAR3,QTMC1},
most notably through reducing friction \cite{friction}
%%%FFF OK
and serving as a fuel \cite{fuel}. %,fuel2}. 
In quantum optics, quantum coherences bring about, e.g., the electromagnetically induced transparency effect \cite{EIT2,EITexp,EIT} and coherent population trapping \cite{CPT,Scully,CPT2}, and are used for lasing without inversion \cite{LWI2,LWI,LWIX,Scullybook,LWI5}. In photovoltaics, coherences can increase efficiency \cite{photovoltaics,DBc2,db1}. Quantum coherences are also essential to quantum sensing \cite{Paola} and metrology \cite{Metro}. 
The goal of this work is to investigate the generation of nonequilibrium quantum coherences, and their manifestation and roles  in steady-state heat transport in pursuit of coherent control of thermal transport in devices, and a potential quantum advantage. % DDD2 delete?

\begin{figure}[hbt!]
\centering
\includegraphics[width=0.95\columnwidth]{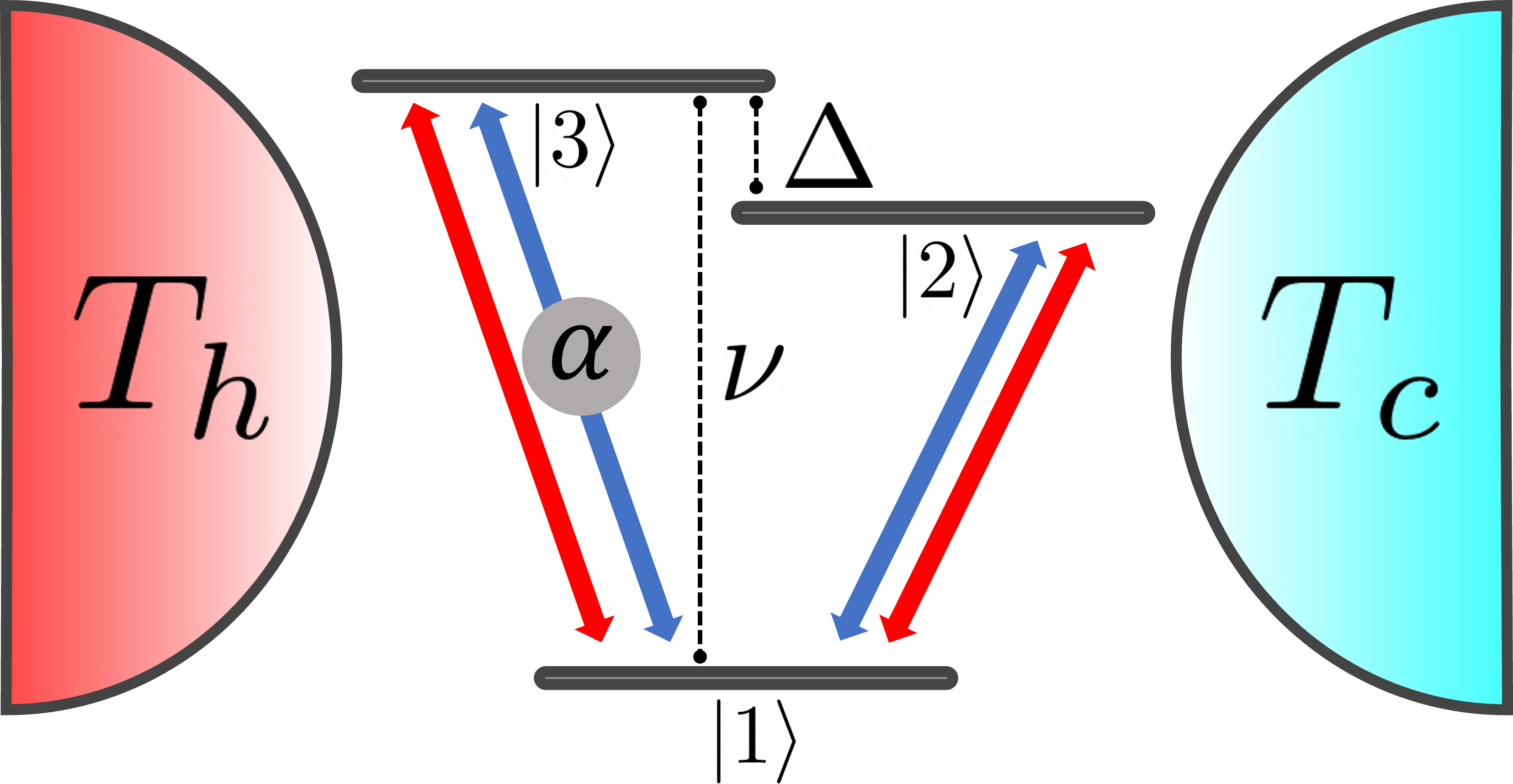} %{Figures/Diagram1new.png}
\caption{Schematic representation of the V model with quasi-degenerate excited states, $\Delta\ll\nu$.
The two heat baths ($T_h> T_c$) allow transitions between the ground state $|1\rangle$  and the two excited states, $|2\rangle$ and $|3\rangle$, 
thus realizing a two-arm interferometer: Heat is transferred from the hot  ($T_h$) to the cold ($T_c$) bath
through two paths, which are distinguishable when  
$|\alpha|\neq 1$, for vanishing $\Delta$.
The sign and magnitude of $\alpha$, a real-valued tunable parameter, determines whether the interference pattern in steady state is perfectly constructive ($\alpha=-1$) or destructive ($\alpha=1$). }
\label{fig:Fig1}
\end{figure}

% V model
The V level scheme is one of the simplest models that can be analyzed in revealing the role of quantum coherences in dynamical and steady state properties.
This model, and the related $\Lambda$ model, have been analyzed in great details in the context of quantum optics \cite{PhysRevA.46.373,PhysRevA.47.2186,Agarwal1999,CPT,Li_2000,Scully,EIT,EITexp,LWI5,AO,KIFFNER201085,Vmodelcorr}, 
and have more recently received renewed attention
%and more recently in quantum biology 
motivated by insights they offer in the study of light-harvesting systems \cite{JCao,Timur14,dodin_quantum_2016,b4,b3,b1,b2}. % DDD REFS 
The V model is composed of three levels: A ground state and two quasi-degenerate excited states, with coherences generated between the two excited levels due to driving by coherent or incoherent environments. Many variants of this model exist; in our work, the ground state is coupled to both excited states through two heat baths, 
see Fig. \ref{fig:Fig1}. Each transition (ground state $\leftrightarrow$ each excited state) leads to heat flow between the hot and cold baths. However, the two transitions (paths) can be tuned to  interfere constructively or destructively with each other.
%some results on V model

Long-lived transient quantum coherences were shown to be generated in the V model when coupled to a {\it single} heat bath, see e.g., Refs. \cite{b1,b2,b3,b4,Timur14,dodin_quantum_2016}. Though of significant magnitude and lifetime, these quantum coherences eventually decay and vanish in the steady-state regime, limiting potential applications in e.g. autonomous quantum heat engines. 
Quantum coherences were demonstrated to be sustained in the steady-state limit once the V model was coupled to nonequilibrium reservoirs or when detailed balance was broken by other means \cite{Agarwal,Scully,Li,VmodelNJP,JCao,VRect,b5,SSC}. 
%%%FFFF our system obeys detailed balance and is in nonequilibrium, but in certain cases e.g. a=1, coherences are zero. May need to clarify? 
% DDD4 rephrased to make it clear that the sentence was about the cited work
In both transient and steady state regimes,  coherences can be generated due to incoherent processes instead of benefiting from coherent sources such as laser or cavity confinement \cite{Obada_2020}, in which case \textit{transfer} of coherence (e.g., from the laser to the system) more appropriately describes the phenomenon. 
Despite this similarity, transient and steady state coherences are controlled by different parameters, as we shall demonstrate in this work. 
%While approaches used in Refs. \cite{b5,Scully,Li,Agarwal} invoke arguably strong approximations e.g., weak coupling to the external environment, results are consistent with thermodynamical laws \cite{Agarwal} and in particular do not utilize decoherence-free-subspaces \cite{Li,Whaley}.
% DDD delete?

%The two phenomena, as we shall also show in this work, are therefore likely rooted in different physics with the nonequilibrium character of the latter accordingly suggesting its bath-induced origin. W
 
%It is known that the near-degenerate states of a three-level V-model builds long-lived (but transient) quantum coherences when equilibrated by the equivalent of an incoherent thermal bath \cite{b1,b2,b3,b4}. They do not rely on coherently-induced transitions (e.g., laser), therefore highlighting the more physical picture of incoherent energy sources such as phonons. Surprisingly, non-vanishing quantum coherences may also generate in the steady-state limit, if the V-model is in a non-equilibrium setting and when certain conditions \cite{b5,Scully,Li,Agarwal,Jcao} are satisfied. While approaches used in Refs. \cite{b5,Scully,Li,Agarwal} invoke arguably strong approximations e.g., weak coupling to the external environment, results are consistent with thermodynamical laws \cite{Agarwal} and in particular do not utilize decoherence-free-subspaces \cite{Li,Whaley}. In what to follow, quantum coherences are understood to mean the non-diagonal elements of the density matrix describing the system in the energy eigenbasis.

% Electrons
In electron transport junctions, steady-state quantum coherences have been explored intensively, both theoretically \cite{QDT6,QDT5,QDT2,QDPRB,QDPRB2,QDT4,QDT3}  and experimentally \cite{QDE1,QDE8,QDE2,QDE3,QDE4,QDE5};  %,QDE6,QDE7}; %,QDE9};
references here are examples of a rich literature. 
In a typical setup, two quantum dots are coupled to two metals through separate arms, a reference arm, and an adjustable arm, constituting
 the realization of an electron interferometer. The phase difference between the two paths
can be tuned by a magnetic flux via the Aharonov–Bohm effect \cite{AB}. %,QDT1}.
 %(the Mach-Zehnder interferometer is the optical equivalent, as suggested in Ref. \cite{Plenio}). 
The type of interference between the two arms controls the extent of quantum coherences between the quantum dots, and thus the steady state charge current.
%The mechanism by which quantum coherences persist in the steady-state limit is through---and probed with---phase interference\cite{Kubo} between the two transition paths, made possible by population constraints at steady-state.
Using the nonsecular Redfield master equation,  which is a \textit{microscopic} technique that treats quantum coherences faithfully, 
here we demonstrate the connection and analogy between the V model and the double-quantum-dot setup.
%and show that the physics underlying steady-state quantum coherences between the two setups is the same. 
%Our approach benefits from extensive generality and, as far as we know, unique. 

% Here
In this work, we use the V model with quasi-degenerate excited states as a thermal-conducting element, mediating heat transport between two thermal baths. 
Since heat is transported through two ``arms", 
coherences develop between the excited states. By controlling the balance between the arms, interference effects are modulated to enhance or suppress the heat current. The V model thus demonstrates
coherence-control of thermal energy transport. 
We describe this effect in both the transient and steady state limits.
%using the  nonsecular Redfield quantum master equation. %While typically cumbersome to be solved, here 
Our key new results are: 

(i) We derive closed-form expressions for coherences, populations and the heat current in the steady state regime using a nonsecular quantum master equation, identifying conditions and the extent of control of heat transport via coherences. 
As we show below, coherences between excited states are nonzero in steady state due to the interplay of coherent and thermal effects.
Their real part satisfies
\bea
\sigma_{32}^R&\propto& 
(\alpha-1)^2\times \left [n_h(\nu)-n_c(\nu)\right] 
\nonumber\\
&\times& \left\{ J_h(\nu)[n_h(\nu)+1 ] - \alpha J_c(\nu)[n_c(\nu)+1 ]\right\},
\label{eq:sscohMN} %FFF OK
\eea
with the imaginary part proportional to the real part.
Here, $\alpha$, a dimensionless real parameter,  controls the asymmetry between the interfering arms. $n_{h,c}(\nu)$ is the Bose-Einstein occupation factor of the $h,c$ baths and $J_{h,c}(\nu)$ is the spectral density function of the heat baths.  Both functions are calculated at the energy gap $\nu$, between ground and excited states.
Similarly, we show that the steady state heat current can be tuned by $\alpha$,
\bea
j_h\propto \nu(\alpha+1)^2 [n_h(\nu)-n_c(\nu)]
\label{eq:jhint}. %FFF OK
\eea
Note that the coherences and the heat current are nonmonotonic in $\alpha$, as we show in Sec. \ref{S:Analytical} with full expressions and simulations. 
%as auxiliary $\alpha$-dependent
%terms are implicit.

(ii) We confirm numerically the thermodynamic consistency of the nonsecular Redfield equation in the present model through an analysis of the dynamical map for the explored range of parameters.

(iii) We demonstrate stark differences between transient dynamics of the V model and its steady state behavior, suggesting that different parameters are at play for either generating or sustaining coherences:
While the energy splitting between the closely-degenerate excited stated dictates the lifetime of transient dynamics, this energy has only a secondary, small role in dictating steady state values.

%Outline
This work is organized as follows. 
In Sec \ref{S:Model}, we describe the V model. In Sec. \ref{S:Method}, we outline the Redfield quantum master equation that we use to study dynamics and steady state transport. 
We present simulations of the transient dynamics in Sec. \ref{S:Transient} and analytic results in the steady state limit
in Sec. \ref{S:Analytical}.
In Sec. \ref{S:Discussion} we discuss (i) the local-site picture and its physical interpretation of quantum coherences, (ii) the thermal diode effect, and (iii) the thermodynamic consistency of our approach. We conclude in Sec. \ref{S:Summary}.

%===============================================================
% Figure 1
%================================================================

\section{Model}
\label{S:Model}

The nonequilibrium V model (Fig. \ref{fig:Fig1}) is a minimal
model that exhibits nontrivial quantum coherences in its operation as a thermal junction, at weak system-bath coupling.
In the energy basis, the Hamiltonian of the model is given by 
\bea
\hat{H}_S=(\nu-\Delta)|2\rangle\langle2|+\nu|3\rangle\langle3|.
\label{eq:Hs}
\eea %%%FFF OK
We adopt here natural units, $\hbar=1$, $k_B\equiv1$.
To facilitate interference effects, we work  in the limit of  $\Delta\ll\nu$. 
Transitions between states of the system,
$|1\rangle\leftrightarrow|2\rangle$ and $|1\rangle\leftrightarrow|3\rangle$ are 
enacted by two independent bosonic heat baths denoted by $h$ (hot) and $c$ (cold) with $T_h>T_c$. The Hamiltonian of bath $m$ is
\bea
\hat{H}_{B,m}=
%\sum_{m={h,c}}
\sum_{k}\nu_{k,m}\hat{b}_{k,m}^\dagger\hat{b}_{k,m}.
\eea %%%FFF OK
Here, $\hat{b}_{k,m}^\dagger$ ($\hat{b}_{k,m}$) is the creation (annihilation) bosonic operator of a mode $k$ of frequency $\nu_{k,m}$ in bath $m$.
The system-bath interaction Hamiltonian is given in a bipartite 
form, with a system operator $\hat{S}_m$ coupled to a bath operator $\hat B_m$,
\bea
\hat{V}_m=  \hat{S}_m\otimes\hat{B}_m;   \,\,\,\,\,\
\hat{B}_m=\sum_k\lambda_{k,m}(\hat{b}_{k,m}^\dagger+\hat{b}_{k,m}). 
\label{eq:Vm}
\eea %%%FFF OK
 $\lambda_{k,m}$ describes the system-bath coupling energy between mode $k$ in the $m$th bath and the system.
To control the interference pattern between the two arms, 
we introduce an asymmetry in transition couplings with a real-valued dimensionless parameter $\alpha$, so that
\bea
\hat{S}_h=|1\rangle\langle2|&+&|1\rangle\langle3|+h.c.
\nonumber\\
\hat{S}_c=|1\rangle\langle2|&+&\alpha |1\rangle\langle3|+h.c. 
\label{eq:S} %%%FFF OK
\eea
Here, $h.c.$ is a hermitian conjugate.
%The parameter $\alpha$  serves to control the interference pattern.
%%%FFFF repetition of above
Note that a more generalized form for this coupling would replace $\alpha$ by a complex 
number $\alpha e^{i\theta}$. %  \cite{commentAB}. %FFF? DDD2magnitude and phase
In the spirit of the Aharonov-Bohm interferometer,
$\hat{S}_h$ corresponds to  the reference arm while $\hat{S}_c$ is  the adjustable arm. 
In sum, the total Hamiltonian is given by
\bea
\hat{H}=\hat{H}_S+\sum_{m={h,c}}(\hat{H}_{B,m}+\hat{V}_m).
\label{eq:Htot} %FFF OK
\eea
Numerous studies had approached this system in the energy basis of $\hat H_S$
\cite{b1,b2,b3,b4,b5,Timur14,dodin_quantum_2016,Scully,Li,Agarwal,JCao}. % in Fig. \ref{fig:Fig1}. 
However, as we discuss in Sec. \ref{S:Discussion}, interference effects can also be 
visualized in the local basis, as was done, e.g., in Ref. \cite{MK}. 
%analogous to a double-quantum-dot setup with tunneling interactions. 
%We expand on this picture in Appendix \ref{App:localglobal}.

%=================================================================
\section{Method: Review of the Redfield Equation}
\label{S:Method}

The main objective of this work is to understand how quantum coherences can be used to tune and control 
 heat currents in a minimal model of a quantum heat conductor. This is achieved by 
deriving closed-form expressions for quantum coherences and 
heat transport in the steady state limit. These expressions are expected
to serve as the groundwork for understanding more complex systems (e.g., those amenable to strong coupling effects and non-Markovian dynamics).

To allow analytical work, we contain ourselves with the second-order Markovian quantum master equation of the Redfield form, which is not secular, i.e., population and coherences are coupled \cite{Nitzan}.
The assumptions behind Markovian Redfield equations are that
(i) the system and the baths are weakly coupled thus their states
are separable at any instant (Born approximation), and
(ii) that the baths' time correlation functions, which dictate transition rates in the system, 
quickly decay, thus the dynamics of the system is Markovian. 
The resulting Markovian Redfield equation for the reduced density matrix  (RDM) $\sigma(t)$ in the Schr\"odinger representation is given by
\bea
\dot{\sigma}_{ab}(t) &=& -i\omega_{ab}\sigma_{ab}(t) 
\nonumber\\
&-&\sum_{c,d,m} \Big\{R_{ac,cd}^m(\omega_{dc})\sigma_{db}(t)+R^{m,*}_{bd,dc}(\omega_{cd})\sigma_{ac}(t)
 \nonumber \\
&-&[R^m_{db,ac}(\omega_{ca})+R^{m,*}_{ca,bd}(\omega_{db})]\sigma_{cd}(t)\Big\}.
\label{eq:Redfield} %%%FFF OK
\eea
Here, $\omega_{ab}=E_a-E_b$ are the Bohr frequencies with $E_{a}$ being an eigenenergy of the system, see Eq. (\ref{eq:Hs}). 
The index $m$ identifies the bath at issue (cold, hot). 
The Redfield dynamics can be also written in a compact form as 
$\dot \sigma(t) = -i[\hat H_S,\sigma(t)] +\mathcal{D}_h(\sigma(t)) + \mathcal{D}_c(\sigma(t))$,
with the dissipators $\mathcal{D}_m(\sigma(t))$.
Terms in the dissipators are given by half Fourier transforms of autocorrelation functions of the baths,
\bea
%R_{ab,cd}^j(\omega)&=&S_{ab}^j S_{cd}^j \int_{0}^{\infty} d\tau e^{i\omega\tau}\langle \hat{B}_I^j(\tau)\hat{B}_I^j \rangle \nonumber\\&=&S_{ab}^j S_{cd}^j [\frac{k^j(\omega)}{2} + i D^j(\omega)] \label{hft}
R_{ab,cd}^m(\omega)&=&(S_{m})_{ab} (S_{m})_{cd} \int_{0}^{\infty} d\tau e^{i\omega\tau}\langle \hat{B}_{m,I}(\tau)\hat{B}_{m,I} \rangle \nonumber\\&=&(S_{m})_{ab} (S_{m})_{cd} 
\left[\frac{k_m(\omega)}{2} + i Z_m(\omega)\right].
%\nonumber\\&=& [{\Gamma^j(\omega)} + i D^j(\omega)]\label{hft}
\label{eq:Rabcd} %%FFFF moved m superscript down to be consistent with Eq. 5 (also superscript)
\eea
Here, operators are given in their interaction representation with respect to $\hat H_S+\sum_m\hat H_{B,m}$.
The matrix elements $(S_{m})_{ab}$ are given in Eq. (\ref{eq:S}).
The real and imaginary parts of the correlation functions can be written in terms of the spectral density function of the bath, $J_m(\omega) = \pi\sum_k\lambda^2_{k,m} \delta(\omega-\nu_{k,m})$.
%so that for bath $\alpha$:
%In this work, the imaginary terms will be neglected. There are, admittedly, ongoing research as to if this so-called secular approximation is well justified, especially in coherent systems -- regardless, a hand-wavy argument I will plead to is the following: the dynamical component attributed to the tensor $D_{ij,kl}(\omega)$, responsible for the imaginary part of $R$, is oscillatory and when it is 'fast' it averages out.
For the interaction model Eq. (\ref{eq:Vm}), the real part $\frac{k_m(\omega)}{2}$  reduces to
\bea 
\label{eq:rate}
\frac{k_{{m}}(\omega)}{2} =
\begin{cases}
 J_{{m}}(\omega) n_{m}(|\omega|) & \omega < 0   \\
 J_{{m}}(\omega)[n_{m}(\omega) + 1] & \omega > 0. 
\end{cases} %%%FFF OK
\eea
As for the imaginary part, $Z_m(\omega)$, we neglect it in this work assuming that it is small at high enough temperatures.  % DDD2 check papers by Juzar

The expressions that we derive below do not rely on a specific form
for the spectral density function, $J_m(\omega)$. In numerical simulations
we assume Ohmic spectral density functions with an infinite high frequency cutoff,  $J_m(\omega) = \gamma_m\omega$ with $\gamma_m$ a dimensionless coupling parameter taken identical for both baths.
The occupation function of the baths follows the Bose-Einstein distribution,
$n_m(\omega)\equiv[e^{\beta_m \omega}-1]^{-1}$ with
$\beta_m=1/T_m$ as the inverse temperature.
Locally, the transition rates obey the detailed balance relation,  
$k_m(\omega)=e^{\beta_m\omega}k_m(-\omega)$.

To calculate heat currents, we study the energy of the system, $ \langle \hat{H}_S (t) \rangle \equiv  {\rm Tr}_S[\sigma(t) \hat H_S]$. 
Using the dynamical equation for the RDM, the rate of energy change of the system is
\bea
\frac{d\langle \hat{H}_S\rangle}{dt}
= {\rm Tr}_S\left[\hat{H_S} \sum_m\mathcal{D}_m\big({\sigma(t)}\big)\right].
\label{eq:curex}
\eea %%%FFF OK
Since the dissipators are additive, %(given the weak coupling approximation),
we identify the heat current flowing into the system from the hot bath as
$j_h(t)= {\rm Tr}_S\left[\hat{H}_S \mathcal{D}_h\big({\sigma(t)}\big)\right]$.
A similar definition holds for $j_c(t)$.
In the long time limit,
$j_h= {\rm Tr}_S\left[\hat{H}_S \mathcal{D}_h\big({\sigma}\big)\right]$ and
$j_h=-j_c$; the heat current is defined positive when flowing from the $m$th bath towards the system. 
Note that we suppress the time variable  to indicate steady state quantities.
To evaluate the current in the steady state limit, 
we first solve the algebraic equations for the RDM by setting $\dot{\sigma} = 0$ with the 
normalization condition $\Tr[\sigma] = 1$, then substitute the results into the dissipator.
%FFFF repetition
%,to calculate the heat current.

In what follows, we study the behavior of the system in both the transient regime and the steady state limit, with and without coherences:
(i) We solve numerically Eq. (\ref{eq:Redfield}) and gather 
both the dynamics of the system and its steady state behavior. 
%These equations are non-secular, thus showing the role of coehrences on the heat current.
(ii) In the steady state limit, we simplify the nonsecular Redfield equation and derive 
closed-form expressions for $\sigma$ and $j_h$ in the limit of small, yet nonzero $\Delta$. 
(iii) We further make the secular approximation and obtain analytic results in the steady state limit. 
Comparisons between (i) and (ii) illustrate fundamental differences between transient dynamics and steady state behavior.
Comparing results from (ii) to (iii) reveals the role of coherences on steady state heat transport in the V model.

%================
% Figure 2
\begin{figure*}[htbp]
\centering
\includegraphics[width=2\columnwidth]{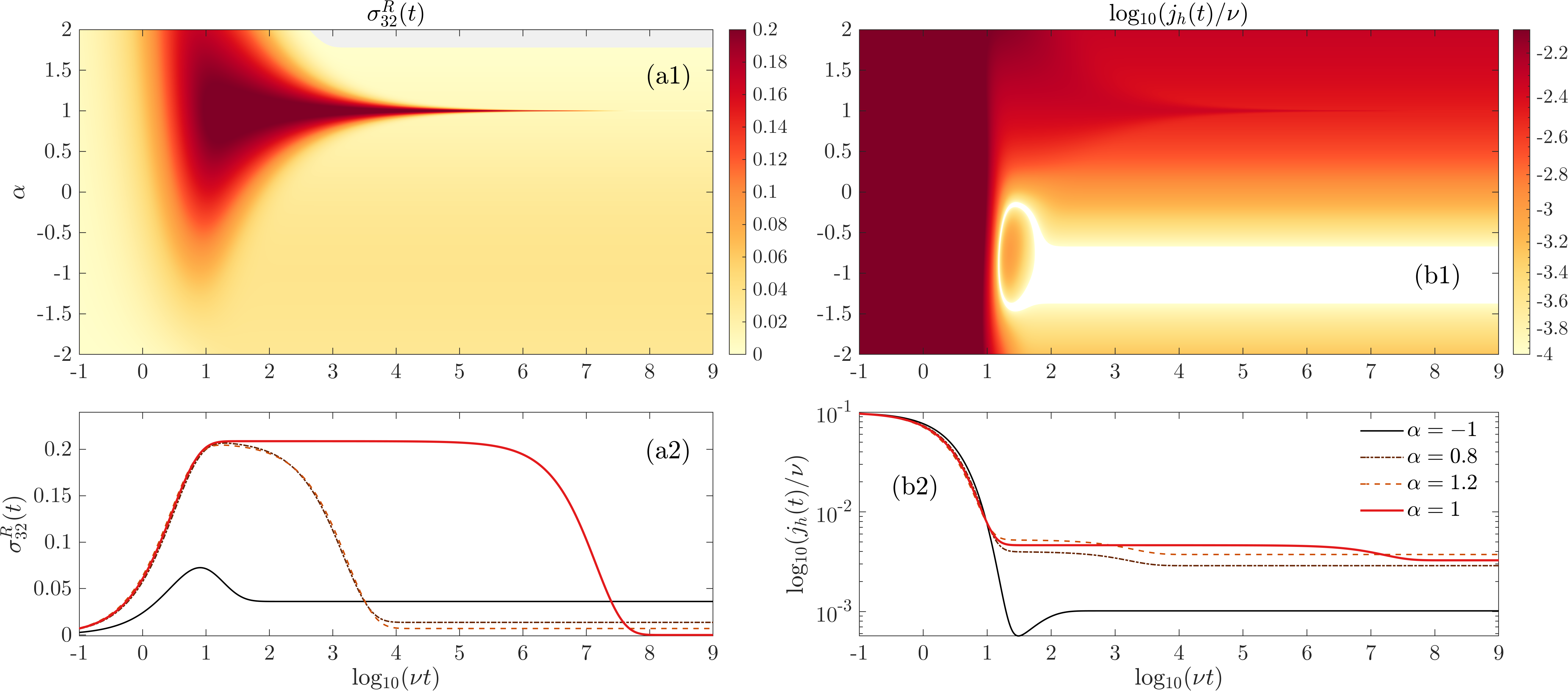}
%%%FFFF is 2.2columnwidth intentional? DDD4 - fixed
%{Figures/Newfig2.png}
\caption{Transient behavior of the nonequilibrium V model. 
% plotted as a function of time and the control parameter $\alpha$.
%Contour plot of the 
(a1) Contour plot and (a2) sections of the real part of quantum coherences as a function of time and the control parameter
$\alpha$. % shown as a contour plot (top) and
 (b1) Contour plot and (b2) sections of the heat currents flowing from the hot bath.
 %to the cold bath. FFFF
%$j_h(t)$ (b1) with slices (b2) as a function of time and the parameter $\alpha$.
The grey area in panel (a1) indicates negative values of the real part of quantum coherences.
%small values (below 0.01) that cannot be smoothly represented.
%%%FFF grey is negative
%by the chosen color scheme.
%the colorbar.
%
Physical parameters are  $T_h=4$, $T_c=2$, $\Delta=10^{-4}$, $\gamma=0.0071$; parameters are given relative to $\nu=1$.}
\label{fig:Cohdyn}
\end{figure*}

%==============================
% Figure 3
\begin{figure}[htbp]
\centering
\includegraphics[width=0.95\columnwidth]{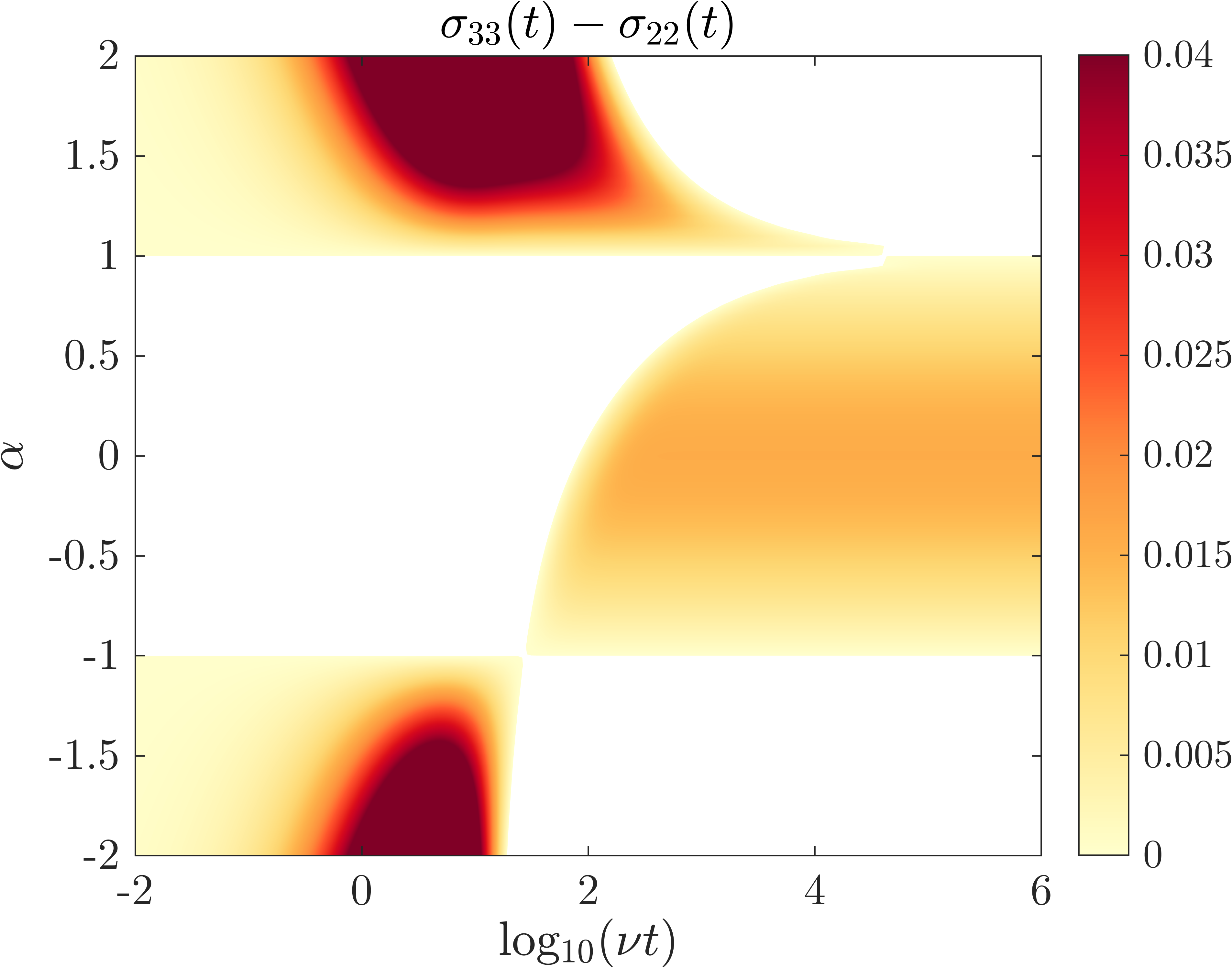}%{Figures/Newfig3.png}
\caption{Temporal behavior of the population difference
$\sigma_{33}(t)-\sigma_{22}(t)$ as a function of $\alpha$. 
White coloring indicates $\sigma_{33}(t)<\sigma_{22}(t)$ (no population inversion). Colored regions represent 
population inversion, which occurs in the long time limit when $|\alpha|<1$.
Parameters are the same as in Fig. \ref{fig:Cohdyn}.}
%$T_h=1.5$, $T_c=0.5$, $\Delta=10^{-4}<<\nu=1, \gamma=0.0071$. $\hbar=k_B\equiv1.$}
\label{fig:Popdyn}
\end{figure}
%=------------------------------------------------------
%========================

\section{Transient Dynamics} 
\label{S:Transient}

We adopt the Redfield equation (\ref{eq:Redfield}) to simulate the system's dynamics; Appendix A
%%%FFFF no reference with hyperlinks? % DDD4: It generated a Sec. index.
includes the specific equations of motion for the V model.
As the initial condition, we assume that only the ground state is populated $\sigma_{11}(t=0)=1$ while all other elements of the RDM are zero. However, we confirmed with simulations that observations were generic for other initial conditions.
(We are not considering here exotic initial conditions that lead to anomalous open-system relaxation as in the Mpemba effect \cite{Mpemba}). %%%FFFFF unique->exotic
%%%FFFF except mpemba initial conditions, where transient coherences decay rapidly (add?) % DDD4 added
Results are presented in Fig. \ref{fig:Cohdyn}, where we study $\sigma_{32}^R(t)$ and the heat current $j_h(t)$ from the initial time to the steady state limit.

First, in Fig. \ref{fig:Cohdyn}(a), we focus on the behavior of coherences between the excited states. We note an intriguing contrast between the transient regime and the steady state limit:
When transient coherences are long-lived at $\alpha=1$,
they eventually vanish in the long-time limit.
In contrast, for other values of $\alpha$ such as for $\alpha=-1$, 
temporal coherences are smaller in magnitude compared to for $\alpha=1$, 
yet coherences survive in the steady state regime.
Long-lived  temporal coherences were explored in e.g. Ref. \cite{b1,b2,b3,b4,Timur14,dodin_quantum_2016}, %DDD3 do not include b5. 
yet under equilibrium (single bath) conditions.
As such, these studies could not observe the curious contrast shown here, between temporal dynamics and steady-state behavior.
% DDD other initial conditions?
%Understanding the physics underlying this observation is a topic for future work.

The temporal behavior of the heat current is presented in Fig. \ref{fig:Cohdyn}(b). We observe the following:
(i) Long-lived coherences temporarily support higher currents,  which decay to their steady state values only when the coherences are suppressed. This effect is most clearly demonstrated for $\alpha=1$, but it is still visible for other values, $\alpha=0.8, 1.2$.
(ii) As for the steady state behavior: 
The current diminishes to a small value for $\alpha=-1$. In fact, in Sec. \ref{S:Discussion} we show that for $\alpha=-1$
the current scales as $\Delta^2$, thus  when the levels approach degeneracy, $j_h\to 0$.

We continue and study in Fig. \ref{fig:Popdyn} the excited state populations,  focusing on the population difference
$\sigma_{33}(t)-\sigma_{22}(t)$. Since $E_3>E_2$, a positive difference indicates population inversion.
Interestingly, we once more observe contrasting behavior between the transient and the steady state regimes:
 For $|\alpha|\leq 1$ population inversion appears only in the steady state limit, but not in the transient regime.
 In contrast, for $|\alpha|>1$, population inversion develops at short time, but disappears in the long time limit.

It was shown for the V model in equilibrium that
the lifetime of transient effects scales inversely with 
$\Delta^2$, see e.g., Refs. \cite{b1,b2,b3,b4,Timur14,dodin_quantum_2016}. As we show next,  in contrast, $\Delta$ has a marginal 
%yet unignorable % DDD4 next sentence covers it
role on steady state values. Regardless, one must include it for properly deriving results.

%========================================

\section{Analytic results in steady state} 
%for $\Delta \to 0$} % DDD4 We know that it is still important to keep Delta neq 0 in the analysis, so I prefer not to have the Delta to 0 in the title.
\label{S:Analytical}

The Redfield equation for the V model (Appendix A)
is cumbersome to solve analytically---even in the steady state limit.
To make progress, we simplify it as follows:
Since $\Delta \ll \nu$, it is reasonable to assume that
the decay rates from either excited states to the common ground state, activated by the
$m$th heat bath, are about equal,
$k_m(\omega_{21})\sim k_m(\omega_{31})$, %denoted by $k^m$ and 
evaluated at $\nu$.
An analogous  approximation holds for the excitation rates. Recall that since $ \nu>0$, the decay rate is given by
% DDD3
\bea
k_m(\nu) =2 J_m( \nu) \left[ n_m( \nu) +1 \right],
\label{eq:kmnu} %%%FFF OK
\eea
see equation (\ref{eq:rate}). 
To simplify the notation, henceforth we withhold the frequency value and use the short notation $k_m$ to denote the relaxation rate induced by the $m$th bath. Excitation rates are written using the detailed balance relation.
The resulting equations for the RDM, together with the condition of conservation of population, are
\bea
\dot{\sigma}_{32}(t)&=& -i\Delta\sigma_{32}(t)
-\frac{1}{2} \left[2k_h + (1+\alpha^2) k_c\right] \sigma_{32}(t)
\nonumber\\%%% FFF OK
&-& \frac{1}{2}\left( k_h+ \alpha k_c\right) \left[ \sigma_{22}(t)+  \sigma_{33}(t) \right]
\nonumber\\
&+& \left(  k_he^{-\beta_h\nu} + \alpha k_c e^{-\beta_c\nu}  \right)\sigma_{11}(t)
\label{eq:s23} %%% FFF OK
\\
\dot{\sigma}_{22}(t)&=&
-\left(k_h+ k_c\right) \sigma_{22}(t) 
+\left( k_he^{-\beta_h\nu}+ k_ce^{-\beta_c \nu}\right) \sigma_{11}(t)
\nonumber\\
&-&\left(  k_h +\alpha k_c\right) \sigma_{32}^R(t)
\label{eq:s22} %%% FFF OK
\\
\dot{\sigma}_{33}(t)&=&
-\left( k_h+ \alpha^2k_c\right) \sigma_{33}(t) 
+\left( k_he^{-\beta_h\nu}+ \alpha^2k_ce^{-\beta_c \nu}\right) \sigma_{11}(t)
\nonumber\\
&-&\left(  k_h +\alpha k_c\right) \sigma_{32}^R(t).
% DDD2 pls keep ( ) brackets
\label{eq:s33} %%% FFF OK
\eea
Here, $\sigma_{32}^R(t)$ corresponds to the real part of coherences.
For details, see Appendix A.
In the long time limit,
this set of equations can be solved analytically as an algebraic system by setting $\dot{\sigma}=0$ together with the population normalization condition.

%================================
% Figure 5
\begin{figure} [bp]
\centering
\includegraphics[width=0.95\columnwidth]{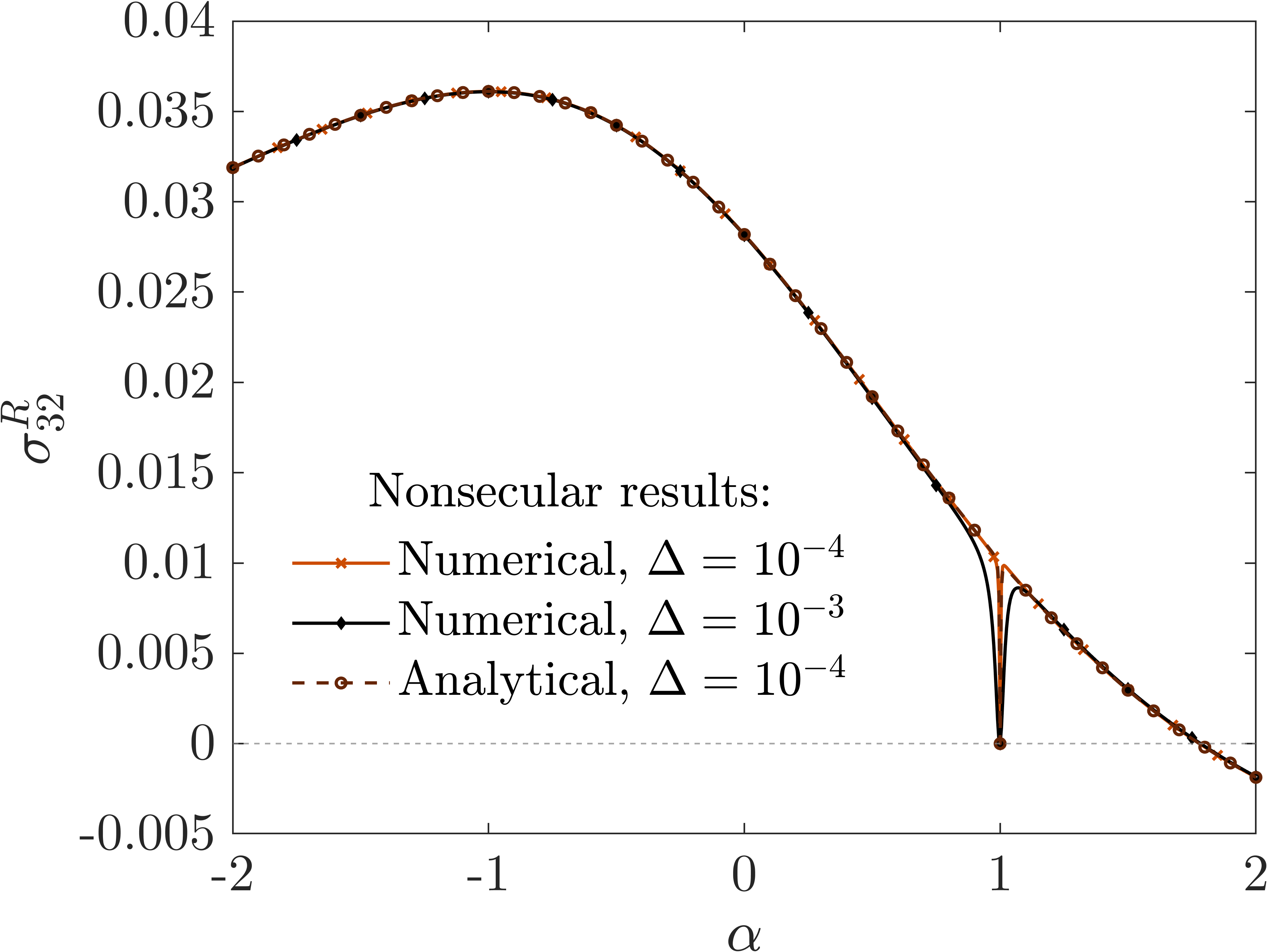} %Fig6new.png}
\caption{
Steady-state quantum coherences calculated numerically
with $\Delta=10^{-4}$ ($\times$) and  $\Delta=10^{-3}$ ($\diamond$),
and analytically using Eq. (\ref{eq:sscohM}) %eq:fsscoh})  % DDD
with $\Delta=10^{-4}$ ($\circ$).
 %Other parameters are $T_h=1.5$, $T_c=0.5$, $\Delta=10^{-4}<<\nu=1, \gamma=0.0071$. $\hbar=k_B\equiv1.$
Parameters are the same as in Fig. \ref{fig:Cohdyn}.}
\label{fig:Fig5}
\end{figure}
%=======================

%================================
\subsection{Quantum Coherences}
In the steady state limit, we find from Eq. (\ref{eq:s23}) that the real ($R$) and imaginary ($I$) parts of the coherences are related via
 %For more details, see \ref{App-EOM}.
\bea
-\Delta\sigma_{32}^{R}=
\underbrace{ \left[\frac{(\alpha^2+1)}{2} k_c+k_h\right]}_{\xi} \sigma_{32}^{I}.
%&+&\frac{k_c\alpha\text{sin}(\theta)}{2}[(1+2e^{-\beta_c\nu})(\sigma_{33}^{ss}+\sigma_{22}^{ss})-2e^{-\beta_c\nu})] 
\label{eq:sscohIR} %FFF OK
\eea
Therefore, taking the real part of Eq. (\ref{eq:s23}) at steady state, we get
\bea % DDD2 use () brackets, not [] unless necessary.
\left( \frac{\Delta^2}{\xi} + \xi \right)\sigma_{32}^R&=&
- \frac{1}{2}\left( k_h+ \alpha k_c\right) \left( \sigma_{22}+  \sigma_{33} \right)
\nonumber\\
&+& \left( k_he^{-\beta_h\nu} + \alpha k_c e^{-\beta_c\nu}  \right)\sigma_{11}.
\label{eq:s23ss} %FFF ok
\eea
Recall that $\sigma$ without an explicit time dependence refers to the density operator in the steady state limit.

Equations (\ref{eq:s22}) and (\ref{eq:s33}) in the long time limit, together with Eq. (\ref{eq:s23ss}) 
and the population normalization condition
provide a closed-form expression for quantum coherences between the two excited states,
%
%\begin{widetext}
\bea
\sigma_{32}^{R} = 
\frac{(e^{-\beta_h \nu }-e^{-\beta_c \nu })  (k_h-\alpha k_c) (\alpha-1)^2} 
{\big[ (\alpha-1)^2 ((\alpha^2+1)k_c+2k_h) (e^{-\beta_c \nu } + e^{-\beta_h \nu } + 1) \big] + \Psi \Delta^2}.
\nonumber\\
 \label{eq:sscohM} %FFF OK
\eea
Here, %$\xi \equiv \frac{2}{2k_h+(1+\alpha^2) k_c}$, and
\bea
\Psi =&& 2\xi(k_c k_h + 2e^{-\beta_h \nu }k_h^2 + k_h^2 + \alpha^2 k_c^2 + 2\alpha^2 e^{-\beta_c \nu }k_c^2
\nonumber \\
&&+ e^{-\beta_c \nu }k_c k_h + e^{-\beta_h \nu } k_c k_h + \alpha^2 k_c k_h 
\nonumber \\
&&+ \alpha^2 e^{-\beta_c \nu }k_c k_h + \alpha^2 e^{-\beta_h \nu }  k_c k_h)/(k_hk_c). %FFF OK
%\nonumber\\
\eea
%
% DDD3 new
Equation (\ref{eq:sscohM}) can be written using the microscopic expression for the rates, $k_m=2J_m(\nu)[n_m(\nu)+1]$, resulting in Eq. (\ref{eq:sscohMN}).

Equation (\ref{eq:sscohM}), or in its microscopic form, Eq. (\ref{eq:sscohMN}), is the first main result of this work. 
It reveals that quantum coherences are non-vanishing in the steady-state limit if the following three conditions are met:

(i) Quantum interferences are nondestructive, with $\alpha\neq1$.

(ii) The system is at a nonequilibrium steady state, $T_h\neq T_c$.
% that is, the baths must be held at different temperatures
%, clarifying the non-equilibrium origin of steady-state quantum coherences

(iii) $k_h \neq \alpha k_c$. 
The equality defines a special point in parameter space.
At this point, the hot and cold bath-induced rates, between the ground state and level $|3\rangle$, are equal, defining a local equilibrium condition. 
This nontrivial special point can be reached either by controlling the temperatures of the baths and their spectral properties, or by tuning the strength of the adjustable ``arm" through the $\alpha$ parameter.

Thus, to maintain steady state coherences, conditions (i) and (ii) separately address the state of the baths (out of equilibrium) and the interference pathway ($\alpha \neq 1$). In contrast, condition (iii) requires that the two aspects, $\alpha$-control and the out-of-equilibrium setting dot not collectively compensate each other and create a local equilibrium situation. 

Numerical results are presented in Fig. \ref{fig:Fig5}. Destructive interferences take place at $\alpha=1$ reflected by a dip
whose width is controlled by $\Delta$. When $\Delta$ vanishes, the $(\alpha-1)^2$ factor cancels and 
condition (i) is obscured. \
Moreover, steady-state quantum coherences displayed by \textit{fully} degenerate V-systems may arise through a distinct mechanism, interpreted as transient coherences that never decay, see Refs. \cite{b1,b2,b3,b4,Timur14,dodin_quantum_2016}, even in equilibrium settings.
%%%FFFF added above Delta=0 implies decay lifetime is infinite (different kind of ss coherences)
It is therefore
critical to perform the analysis at nonzero $\Delta$ to correctly capture the interference behavior around $\alpha=1$.
% Indeed, the regime of $\alpha\to 1$ % strong interferences is characterized by the parameter $\Delta$ and
%is characterized by a  dip in the coherences \cite{QDPRB} as we also show  in Fig. \ref{fig:Fig5}. 
Coherences reach a maximum due to a constructive interference effect when $\alpha=-1$. 
The special point $k_h=\alpha k_c$ at which the coherences vanish is arrived here at $\alpha\approx 1.8 $.

%Remarks on existing studies

Our expression Eq. (\ref{eq:sscohM}) demonstrates clearly that
a proper description of nonequilibrium steady state coherences requires the inclusion of all three aspects: (i) Nonequilibrium settings, $T_h\neq T_c$. (ii) Nondegeneracy of the excited states with $\Delta$ small but nonzero. (iii) Asymmetry of the setup, incorporated courtesy of the control parameter $\alpha$ adjusting the arms.
Although prior studies succeeded in deriving closed-form solutions of nonequilibrium steady state coherences, they did not capture the complete solution Eq. (\ref{eq:sscohM}) as some of these aspects were neglected.
For example, asymmetry was not included in Ref. \cite{VmodelNJP}. As a result, their solution for steady state coherences only depended on thermal occupations of the baths and the eigenenergies of the V-system,  missing the conditions for nonvanishing coherences  $\alpha\neq1$ and $\alpha\neq k_h/k_c$.
Ref. \cite{b5} similarly assumed $\alpha=1$, although coherences persisted in steady state through breaking detailed balance via polarized incoherent light.
Other studies assumed full degeneracy of the excited states, see e.g., Ref. \cite{VRect};
% and thus their solution reflects a different kind of coherence, made manifest by setting $\Delta\equiv0$ in the solutions of Refs. \cite{b1,b2,b3,b4,dodin_quantum_2016}.
%In that case, contrary to our analysis here, steady-state coherences would be observed even in an equilibrium setting.  %DDD4 
%
we stress that setting $\Delta=0$ implies that the dip feature near $\alpha=1$, as shown in Eq. (\ref{eq:sscohM}) and Fig. \ref{fig:Fig5}, is missed, and thus one may infer \textit{incorrectly} that steady state nonequilibrium coherences is nonvanishing even in symmetrical couplings. 
The three aspects discussed here: nonequilibrium, nondegeneracy, and asymmetry were  included in Ref. \cite{Li}, but a closed-form solution was not presented there. 
%Furthermore, the incorporation of asymmetry through allowing different transitions to couple to the same bath with \textit{different} spectral densities, adopted in Refs. \cite{Li,VRect}, is clearly unphysical. 
%

%Altogether, one of our main achievements in this work is deriving  Eq. (\ref{eq:sscohM}), bringing guidelines on how to enhance and sustain steady state coherences.
Complementing this subsection, in Appendix B we show that the previously-analyzed equilibrium V model is a limiting case of our equations of motion.  In Appendix C we further look at the special symmetrical point $\alpha=1$, where steady state coherences vanish even away from equilibrium. 
%====================
% Figure 4
\begin{figure*}[ht]
\centering
\includegraphics[width=2\columnwidth]{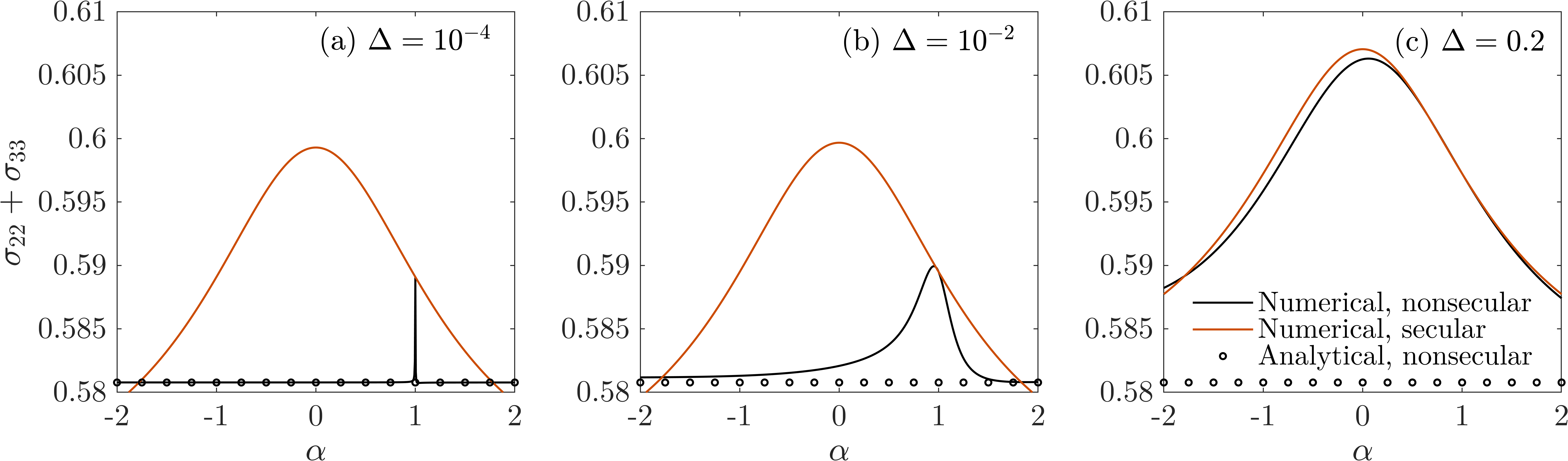} %Fig4new.png}
\caption{Steady state behavior of population.
%%%FFFF Repeated DDD4 made modifications
Plotted is the sum of excited state populations, $\sigma_{33}+\sigma_{22}$,
as a function of $\alpha$ using the nonsecular Redfield analytically (dotted) and numerically (solid black) and secular Redfield numerically (solid grey). In panel (a) for $\Delta=10^{-4}$, numerical and analytical nonsecular Redfield gives the same constant $\sigma_{33}+\sigma_{22}$ as in Eq. (\ref{eq:CPT}) except at the anomalous point $\alpha=1$ where interestingly the nonsecular numerical Redfield result coincides with the secular. The numerical nonsecular results approaches the secular as $\Delta$ is increased, as shown in panel (b) where $\Delta=10^{-2}$ and (c) where $\Delta=0.2$.
    %Other parameters are $T_h=1.5$, $T_c=0.5$, $\Delta=10^{-4}<<\nu=1, \gamma=0.0071$. $\hbar=k_B\equiv1.$
    Other parameters are identical to Fig. \ref{fig:Cohdyn}.}
% DDD?
\label{fig:Popp}
\end{figure*}

%====================
\subsection{Population}

%The steady-state populations are expected to reach values between the Gibbs states of the hot and cold heat baths. 
%Whether such a system evolves to an equilibrium Gibbs state in 
%when coherences are present, have been explored e.g. in Ref. \cite{Agarwal}.
% DDD4 unclear. What is a Gibbs state with coherences?
%Here, we further address questions of thermodynamical consistency in Sec. \ref{S:Discussion}.

%%%FFFF may need to explicitly say that we set Delta=0 here and get solutions bar O(Delta^2)
% DDD4 I think it is OK as is
We now analyze excited state populations in steady state and demonstrate their dependence on the adjustable interference parameter, $\alpha$.
Solving Eqs. (\ref{eq:s22}) and (\ref{eq:s33}) in steady state, together with Eq. (\ref{eq:s23ss}),
we obtain %in the $\Delta\to 0$ limit
\bea
 \sigma_{22}=\frac{ k_c( \alpha^2e^{-\beta_h \nu } + e^{-\beta_c \nu }) 
+ k_h( e^{-\beta_c \nu } + e^{-\beta_h \nu }) }
{\left[(\alpha^2+1)k_c+2k_h) (e^{-\beta_c \nu } + e^{-\beta_h \nu } + 1)\right]} + \mathcal{O}(\Delta^2),
  \nonumber\\\label{eq:fss2v2} %%%FFF OK
\eea
and
\bea
\sigma_{33}=
\frac{ k_c( e^{-\beta_h \nu }  + \alpha^2e^{-\beta_c \nu }) + k_h( e^{-\beta_c \nu } + e^{-\beta_h \nu }) }
{ \left[(\alpha^2+1)k_c+2k_h) (e^{-\beta_c \nu } + e^{-\beta_h \nu } + 1)\right]}+ \mathcal{O}(\Delta^2). 
\nonumber\\ 
\label{eq:fss3v2} %%%FFF OK
\eea
The steady-state populations, $\sigma_{22}$, $\sigma_{33}$, and thus $\sigma_{11}$ 
are insensitive to the sign of $\alpha$. 
% thus to interferences to leading order in $\Delta^2$. 
Furthermore, adding equations (\ref{eq:fss2v2}) and (\ref{eq:fss3v2}) reveals an analog to the 
so-called ``population-locked-states" effect  discussed e.g. in Ref. \cite{Scully},
\bea
%1-\sigma_{11}^{ss}=
P_{ex}\equiv \sigma_{22}+\sigma_{33}=\frac{e^{-\beta_c \nu } + e^{-\beta_h \nu }}{e^{-\beta_c \nu } + e^{-\beta_h \nu } + 1}+ \mathcal{O}(\Delta^2).
 \label{eq:CPT} %FFF OK
\eea
A fraction of the population,
given by Eq. (\ref{eq:CPT}), is always shared between 
the two excited states, $|2\rangle$ and $|3\rangle$, independent of $\alpha$.

Complementing this analytical result, which is valid when $\Delta\to 0$,
in Fig. \ref{fig:Popp} we present
numerical simulations for $P_{ex}$ performed at finite $\Delta$.
As $\Delta$ diminishes,
we observe a sharp spike when approaching  $\alpha=1$, 
at a point where strong interferences between two pathways takes place leading to zero coherences.
Interestingly, at this point the total excited state population 
$P_{ex}$ coincides with the prediction of the secular equation.
 This anomaly is otherwise hidden in the $\mathcal{O}(\Delta^2)$ term in Eq. (\ref{eq:CPT}) (the width of the spike depends on $\Delta$). 
As we increase $\Delta$, we find that the prediction of the secular method approaches the nonsecular result, while the analytic result Eq. (\ref{eq:CPT}), which relies on $\Delta\to0$, naturally fails to capture the correct behavior of $P_{ex}$.
 
From Equations (\ref{eq:fss2v2}) and (\ref{eq:fss3v2}) we find that
both excited states are equally populated when $\alpha=1$ or $\alpha=-1$, up to the first order in $\Delta$.
%FFFF added the up to sentence as this may sound misleading and contradictory to the observed spike.
. Therefore, we can write the state the of the nonequilibrium system as the algebraic average of the two thermal states,
\bea
\sigma (\alpha^2=1) =\frac{e^{-\hat{H}_S\beta_{h}}+e^{-\hat{H}_S\beta_{c}}}{\text{Tr}_S[e^{-\hat{H}_S\beta_{h}}+e^{-\hat{H}_S\beta_{c}}]} + \mathcal{C}+\mathcal{O}(\Delta^2). 
\eea %%%FFFF added the \mathcal{O}(\Delta^2) term
Here $\hat{H}_S$ is the system Hamiltonian of the fully degenerate V model; $\mathcal{C}$ contains the off-diagonal terms of the RDM. 
Furthermore, when $T_h\neq T_c$, we conclude from Eqs. (\ref{eq:fss2v2}) and (\ref{eq:fss3v2}) that
\bea
\alpha^2<1\Leftrightarrow\sigma_{22}<\sigma_{33}. %FFF OK
\eea
Thus,  $\alpha$ can be tuned to force the system to exhibit population inversion  at steady state (although the effect is small with our parameters as shown in Fig. \ref{fig:Popdyn}). This observation is another nontrivial result of this work.
When prepared in the ground state, dynamical simulations in Fig. \ref{fig:Popdyn} also show that 
% always occur between the upper states $|2\rangle$ and $|3\rangle$:
there exists a time interval $\tau$ where $\sigma_{22}(\tau)<\sigma_{33}(\tau)$ before reaching $\sigma_{22}>\sigma_{33}$, and vice-versa.
%-------------------

%=============================
% Figure 6
\begin{figure}[ht]
\centering
\includegraphics[width=0.95\columnwidth]{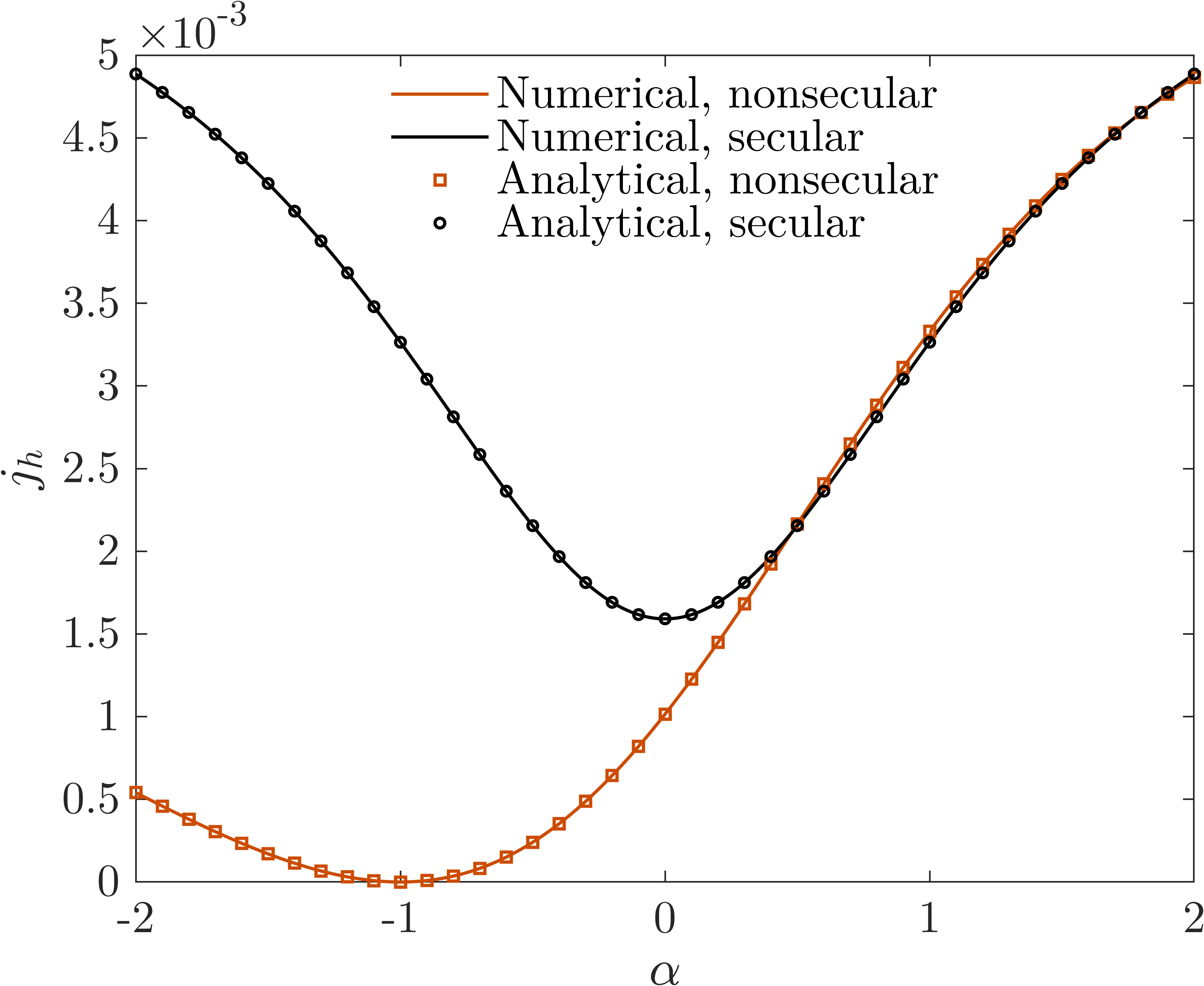} %/Newfig6.png}
\caption{
Steady-state heat current $j_h$ as a function of $\alpha$.
We compare numerical simulations from 
the nonsecular secular (solid black) and Redfield equation (solid orange) to the analytical expressions
 (symbols), with Eqs. (\ref{eq:fsscur}) and (\ref{eq:jhs}), respectively.
 %when $\theta$ is set to 0. Significant differences between both methods manifest when $\alpha<0$. Other parameters are $T_h=1.5$, $T_c=0.5$, $\Delta=10^{-4}<<\nu=1, \gamma=0.0071$. $\hbar=k_B\equiv1.$
Parameters are the same as in Fig. \ref{fig:Cohdyn}.
}
\label{fig:secnonsec}
\end{figure}
%=================================
\subsection{Heat current}
The heat currents $j_{h,c}$ are evaluated at the hot and cold contacts, respectively, using the dissipators, see Eq. (\ref{eq:curex}). Specifically at the hot contact, 
\bea
j_h=\nu\mathcal{D}^{h}_{33}(\sigma)+(\nu-\Delta)\mathcal{D}^{h}_{22}(\sigma).
\label{eq:jhjh} %FFF OK
\eea
Inserting the solutions for the coherences, Eq. (\ref{eq:sscohM}), and the populations, Eqs. (\ref{eq:fss2v2}) and (\ref{eq:fss3v2}) into the dissipator, we get
% DDD3  added nu in front
\bea
j_h=\nu\frac{(1+\alpha)^2(k_hk_c)(e^{-\beta_h \nu }-e^{-\beta_c \nu })}
{\left[(\alpha^2+1)k_c+2k_h\right](e^{-\beta_c \nu } + e^{-\beta_h \nu } + 1)}+
 \mathcal{O}(\Delta).
\label{eq:fsscur} %FFF OK 
\nonumber\\
\eea
% DDD3 OK, the  full expression is not very useful
%
By making the substitution Eq. (\ref{eq:kmnu}) we highlight that the steady state current is non-vanishing when $T_h\neq T_c$ and $\alpha\neq -1$,
and obtain Eq. (\ref{eq:jhint}).
%\bea
%j_h\propto \nu(\alpha+1)^2 [n_h(\nu)-n_c(\nu)]. %FFF %OK
%\eea
%
Note that the $\alpha$-dependence cancels out in Eq. (\ref{eq:fsscur})  when $\alpha k_c=k_h$, a point at which one of the pathways becomes equilibrium-like.

Equation (\ref{eq:fsscur}), or in its microscopic form, Eq. (\ref{eq:jhint}), complements our expressions for the coherences and populations, and it is another central result of this work. We expect that it will be used as the groundwork for more involved studies, e.g., when studying thermal energy transport at strong coupling.
We plot the steady-state heat current in  Fig. \ref{fig:secnonsec}, showing that the above expression is accurate for quasi-degenerate excited states.
Inspecting Eq. (\ref{eq:fsscur}) and Fig. \ref{fig:secnonsec}, we observe the following:  

(i) The nonsecular heat current is {\it not} an even function of $\alpha$, a clear indication that quantum coherences are at play, with the sign of $\alpha$ playing a role in transport. In contrast, under the secular approximation, the current is symmetric in $\alpha$.

(ii) The $[e^{-\beta_h \nu }-e^{-\beta_c \nu }]$ factor in the numerator
exposes the trivial dependence of the heat current on the temperature difference.

(iii) Given the structure of the denominator, it is clear that exchanging the temperatures of the two baths should lead to a thermal diode effect, since the current is not identical under the $k_h\leftrightarrow k_c$ operation (except for $|\alpha|=1$).
%FFFF added the except sentence.
We discuss this aspect  in Sec. \ref{S:Discussion}.

(iv) The steady state heat current is high when quantum coherences are low, and vice versa; compare Fig. \ref{fig:Fig5} to  Fig. \ref{fig:secnonsec}.

(v) To the lowest order, the steady state heat current does not depend on the excited-state splitting $\Delta$. Similarly, the levels' population and their coherences are independent  of $\Delta$ to lowest order  (assuming $\Delta \to 0$). Parameters that do dictate these functions are the spectral density function, temperature of the baths, and the overall splitting $\nu$. In stark contrast, the  rise of coherences and the development of an extended transient region over which they survive is controlled by a timescale inversely proportional to $\Delta^2$, see Refs. \cite{Timur14,b1,b2,b3,b4,b5,comment,dodin_quantum_2016}. % DD3 %%%FFF add footnote on upcoming paper? DDD4 OK
Thus, the lifetime of transient coherence, and their magnitude at steady state are dictated by a different set of parameters. 
%====================================

\subsection{The secular approximation}

The secular approximation decouples population dynamics and coherent dynamics, amounting to crossing out $\sigma_{32}(t)$ in Eqs. (\ref{eq:s22}) and (\ref{eq:s33}).
It is justified when $\Delta$ is large such that the characteristic time of coherent oscillations is much shorter than the population decay time.
%the bath-correlation time. 
%When coherences are finite, we show that the secular approximation nonetheless misses the physics coming from the phase factor term $e^{i\theta}$ in Eq. )\ref{S}).
The steady state solution %of the analogous set of differential equations to Eqs. 
%(\ref{eq:s22}), (\ref{eq:s33}), %(\ref{eq:s23})
thus involves a $2\times2$ coefficient matrix for the population of the excited states.
% (see \Delta^2 \ref{App:EOM} for more details).  % DDD2
At steady state, the heat current is given by ($S$ stands for secular)
\begin{widetext}
\bea
% DDD3 added nu in front
j_h^S = \frac{\nu k_hk_c(e^{-\beta_h\nu}-e^{-\beta_c\nu})\left[k_h(\alpha^2+1)+2\alpha^2k_c\right]}
{(k_h+k_c)(k_h+\alpha^2 k_c)  + e^{-\beta_h\nu} (2k_h^2+k_ck_h+\alpha^2k_ck_h) 
+ e^{-\beta_c\nu} (2\alpha^2k_c^2+k_ck_h+\alpha^2k_ck_h)
}
%(k_c + k_h)(k_c\alpha^2 + k_h)(e^{-\beta_h\nu} + e^{-\beta_c\nu} + 1) + (k_h^2- \alpha^2k_c^2)(e^{-\beta_h\nu} - e^{-\beta_c\nu})
%%FFF which denominator looks better?
+ \mathcal{O}(\Delta). 
\label{eq:jhs} %%%FFF OK
\eea
\end{widetext}
Most notably, this expression is {\it even} in $\alpha$,  unlike the nonsecular result, Eq. (\ref{eq:fsscur}), thus reflecting that only the magnitude of $\alpha$ controls the secular current, rather 
than its sign. This behavior is illustrated in Fig. \ref{fig:secnonsec}.
In the special cases $|\alpha|=1$, Eq. (\ref{eq:jhs}) reduces to
% DDD3 added nu in front
\bea
j_h^S (\alpha^2=1) =
\frac{2\nu k_hk_c(e^{-\beta_h\nu}-e^{-\beta_c\nu})}{k_h(1+2e^{-\beta_h\nu})+k_c(1+2e^{-\beta_c\nu})}+\mathcal{O}(\Delta).\nonumber\\
\label{eq:jhsu} %%%FFF OK
\eea
This expression recovers the result obtained in Ref \cite{MK} where $\alpha=-1$ was used throughout.
% where the secular approximation was indeed shown to have unphysical implications of finite
%current flowing without coupling elements.
Interestingly, Fig. \ref{fig:secnonsec} shows that the secular approximation works reasonably well when  $\alpha\gtrsim1$.
Another nontrivial observation from Fig. \ref{fig:secnonsec} is that for a broad range of parameters, quantum coherences  {\it reduce} the heat current in the V model, with the secular current exceeding the nonsecular one. Coherences are either noninfluential to the current, or they suppress it. This can be viewed as a coherent heat current flowing against the temperature difference   \cite{VmodelNJP}.

%=============================================
% Figure 7
\begin{figure}[htbp]
\centering
\includegraphics[width=0.95\columnwidth]{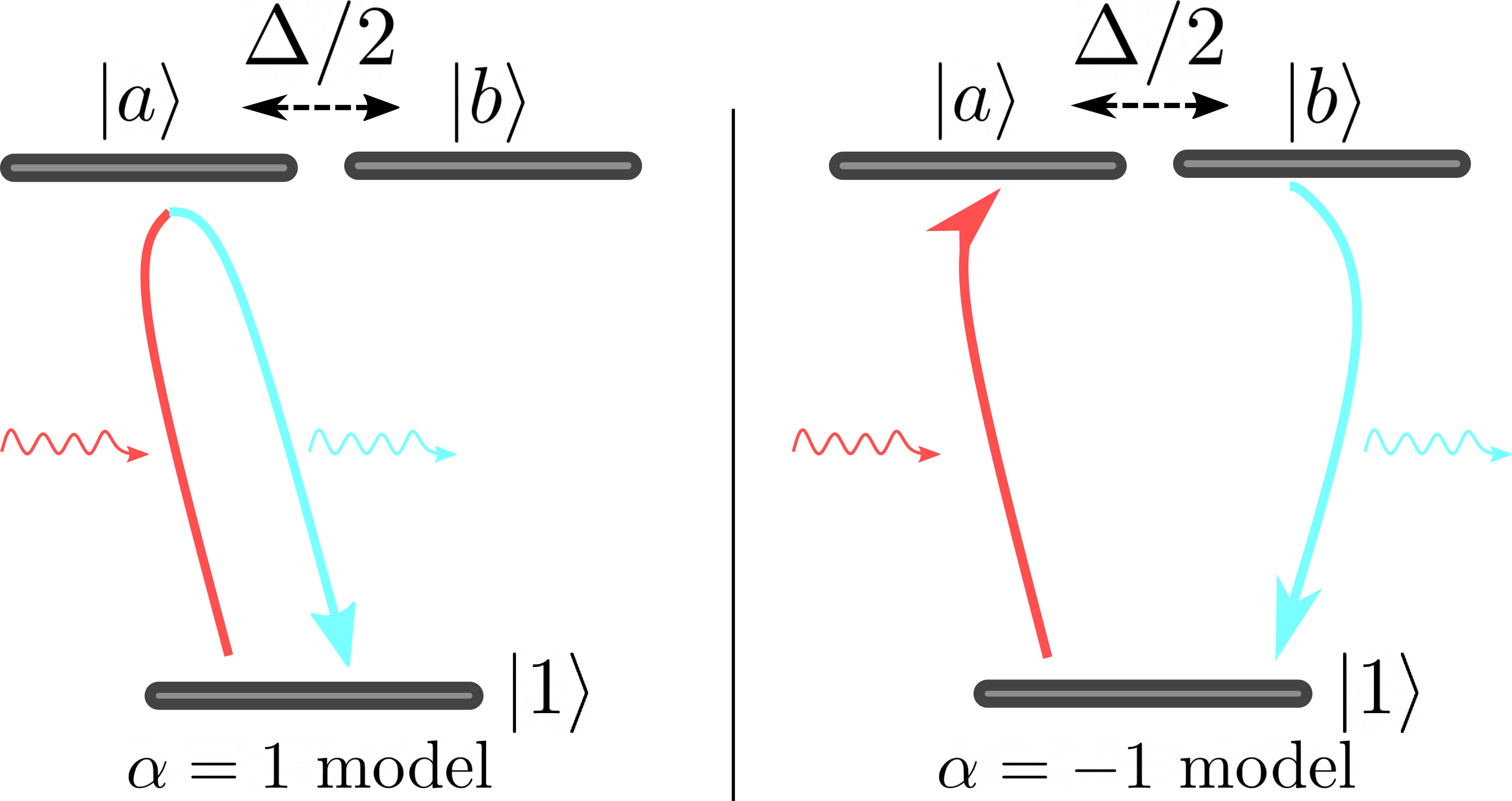} %Fig7.png}
\caption{Schematic representation of the V model in the local site basis. In this picture, $\Delta/2$ is the tunneling energy between degenerate excited states.
(left) The $\alpha=1$ case corresponds to a side-coupled model
where only one transition, $|1\rangle \leftrightarrow |a\rangle$ is coupled to both heat baths, but the excited states are coherently coupled (dashed arrow).
(right) The $\alpha=-1$ case corresponds to a serial model:
Heat is absorbed in the  $|1\rangle \leftrightarrow |a\rangle$ transition,
it tunnels coherently to $|b\rangle$, and decays from there to the cold bath.}
\label{fig:diagramp}
\end{figure}
% 

%------------------
% Figure 8
\begin{figure}[htbp]
    \centering
\includegraphics[width=0.95\columnwidth]{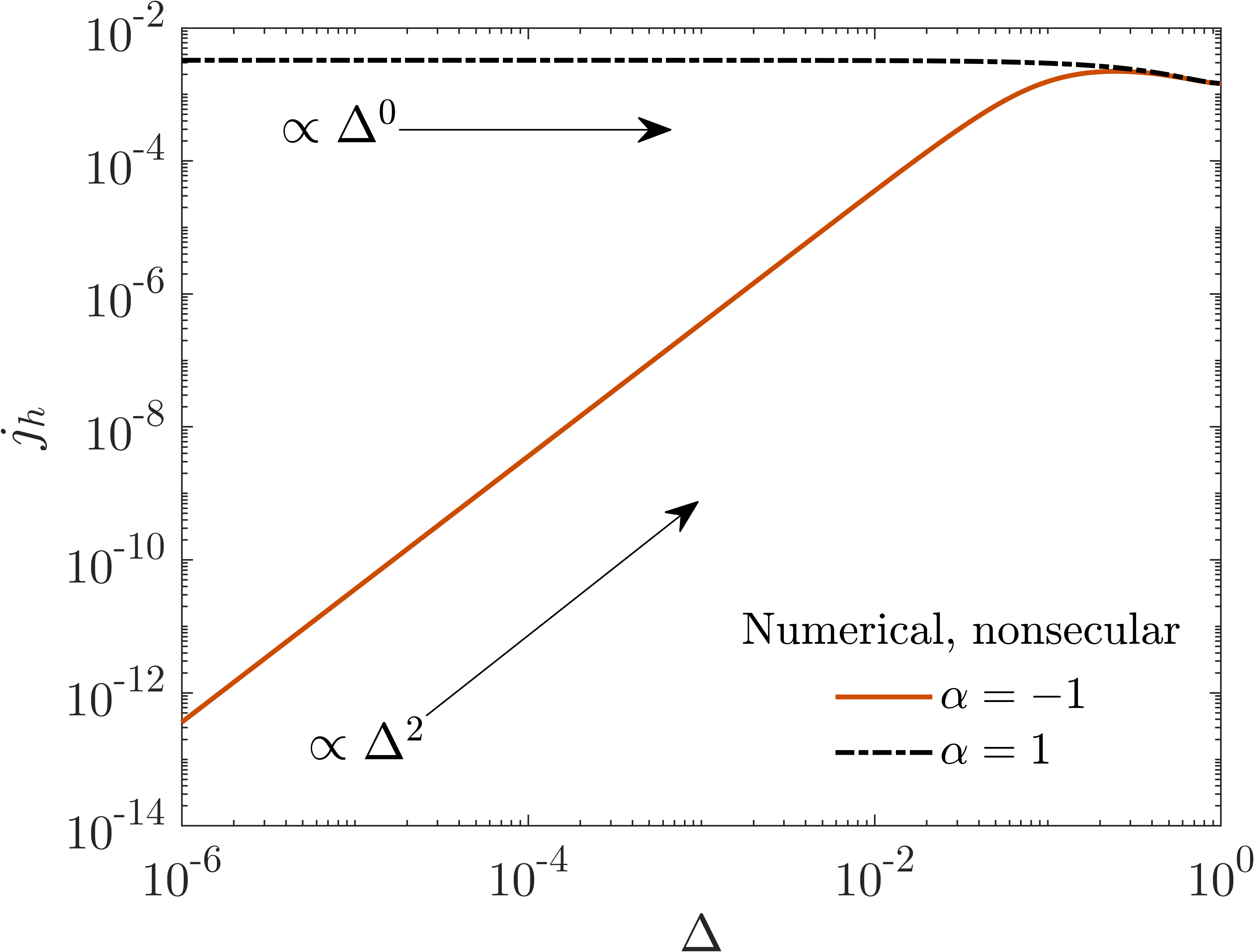} %Fig8new.png}
\caption{Steady-state heat current vs. excited state splitting $\Delta$ 
for $\alpha=-1$ (solid)  and $\alpha=1$ (dashed-dotted).
While in the former, $j_h\propto \Delta ^2$ when $\Delta$ is small, in the latter case of $\alpha=1$ the current is nonzero (and independent of $\Delta$) even when $\Delta \to 0$. 
Other parameters are identical to Fig. \ref{fig:Cohdyn}.}
\label{fig:NP1}
\end{figure}
% ----------
%======================================
%======================================
% Figure 9
\begin{figure*}[htbp]
\centering
\includegraphics[width=2\columnwidth]{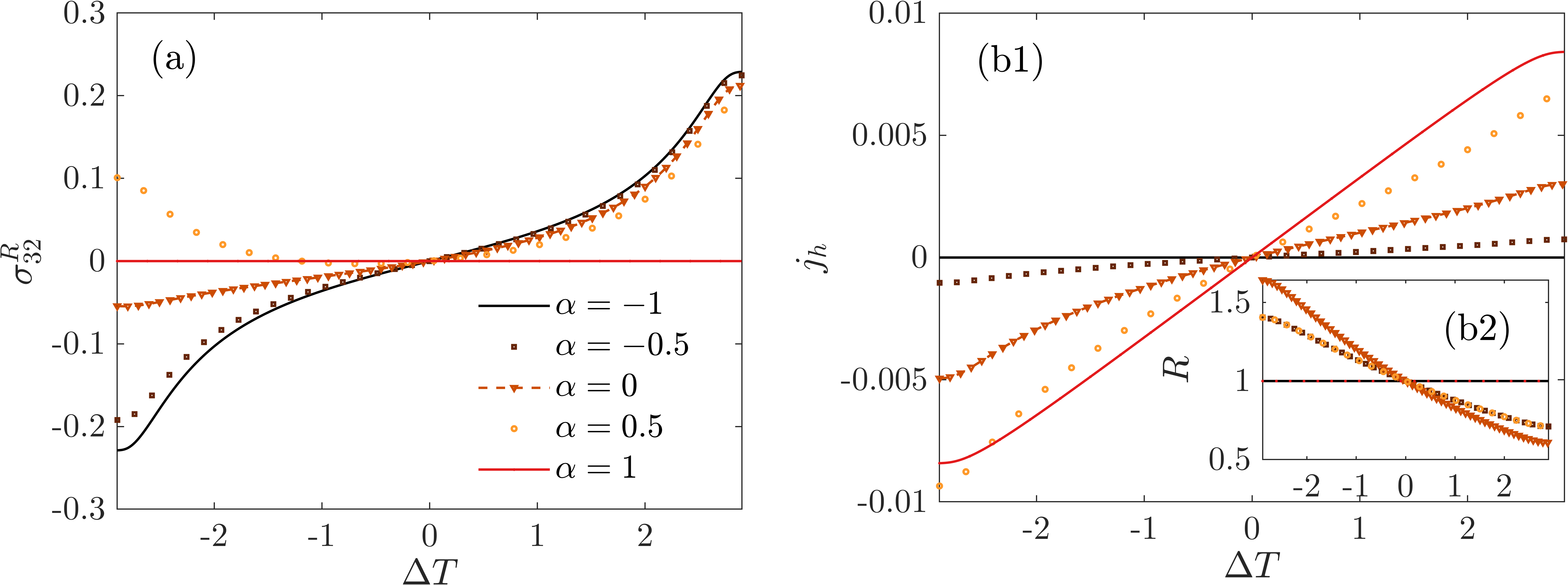} %Fig8new.png}
\caption{Control of thermal rectification by  the coupling strength $\alpha$.
(a) Coherences and (b1) heat current  as a function of the temperature difference $\Delta T=T_h-T_c$.
(b2) The rectification ratio $R=|j_h(\Delta T)/j_h(-\Delta T)|$
for different values of $\alpha$  ranging from $-1$ to $1$.
%Similarly, we observe no dependence to $\Delta$ in the other pathway $\alpha=1$.
% More importantly, this plot shows that heat currents are non-vanishing in this setup, 
%contrary to Fig. \ref{fig:jhcoh} might suggest.  % DDD?
Parameters are the same as in Fig. \ref{fig:Cohdyn}.}
\label{fig:rect}
\end{figure*}
%==============
%------------------
\section{Discussion}
\label{S:Discussion}

\subsection{Quantum interferometers and local-basis mapping}

The V model as depicted in Fig. \ref{fig:Fig1} is  analogous to a double quantum dot  interferometer.
%see e.g., Refs. \cite{QDPRB,QDPRB2}. 
However, instead of electron transport through two parallel dots, here thermal energy transfer takes place through quasi-degenerate 
excited states of frequencies $\nu-\Delta$ and $\nu$.
In our work, the driving force for the transfer process are incoherent baths, rather than a coherent laser, but the mechanism is similar to Fano interferences \cite{Fano}, also considered in Refs. \cite{db1,db2,DBc2,b1,b2,b3,b4,Timur14,dodin_quantum_2016,b5}. 

The effects observed in this work can be also rationalized without 
requiring interference arguments, by working in the local basis \cite{MK}, referred to as the V-Energy Transfer System (VETS). 
The V model, Fig. \ref{fig:Fig1}, can be unitarily transformed into a local (L) picture,
\bea
\hat{H}_S^L=
\left(\nu-\frac{\Delta}{2}\right)(|a\rangle\langle a|+|b\rangle\langle b|)+\frac{\Delta}{2}|a\rangle\langle b|+h.c.
\eea
The three levels in this system are the ground state $|1\rangle$ (same as in the energy basis) and the fully degenerate excited states $|a\rangle$ and $|b\rangle$ with the tunneling energy
$\Delta/2$. This system couples to heat baths via 
\bea
\nonumber
\hat{S}_h^L&=&|1\rangle\langle a|+h.c. 
\\ 
\hat{S}_c^L &=& \frac{1+\alpha }{2}|1\rangle\langle a| + 
\frac{\alpha-1 }{2} |1\rangle\langle b|+h.c.,
\eea
%which are the reference arm, and the adjustable arm 
see Appendix D for further details.

The two limiting configurations, $\alpha=1$ and $\alpha=-1$,
are sketched in Fig. \ref{fig:diagramp}.
The $\alpha=1$ case  corresponds to a ``side-coupled" model in the context of e.g. double quantum dots \cite{QDE9}. In this case,
the transition $|1\rangle \leftrightarrow |a\rangle$  is driven by both heat baths, but 
level $|b\rangle$ is accessible only through a coherent coupling from level $|a\rangle$. In contrast, the $\alpha=-1$ case is reminiscent of a serial double dot model.
Here, the ground state is coupled to level $|a\rangle$ through the hot heat bath.
Since the two excited states are coherently coupled, energy transfers coherently from $|a\rangle$ to level $|b\rangle$,
followed by a relaxation to the ground state and energy transfer to the cold bath using $\hat S_c^L$.
In this picture, it is obvious that the heat current vanishes for $\alpha=-1$ once 
$\Delta=0$.
% DDD2 fixed. Pls check

Consistent with this description, the steady-state heat currents in Fig. \ref{fig:NP1}
demonstrate distinct scaling with $\Delta$ for $\alpha=-1$ and $\alpha=1$:
In the former case,  the current scales 
as $\Delta^2$. On the other hand, the $\alpha=1$ model only 
minimally depends on $\Delta$.
Note that when $\Delta$ is very large (approaching the total gap $\nu$), 
deviations from these scaling are observed due to the large modification of energy levels, see Ref. \cite{MK}.

%=======================================
% Figure 10
\begin{figure*}
[htb]
\centering
\includegraphics[width=2\columnwidth]{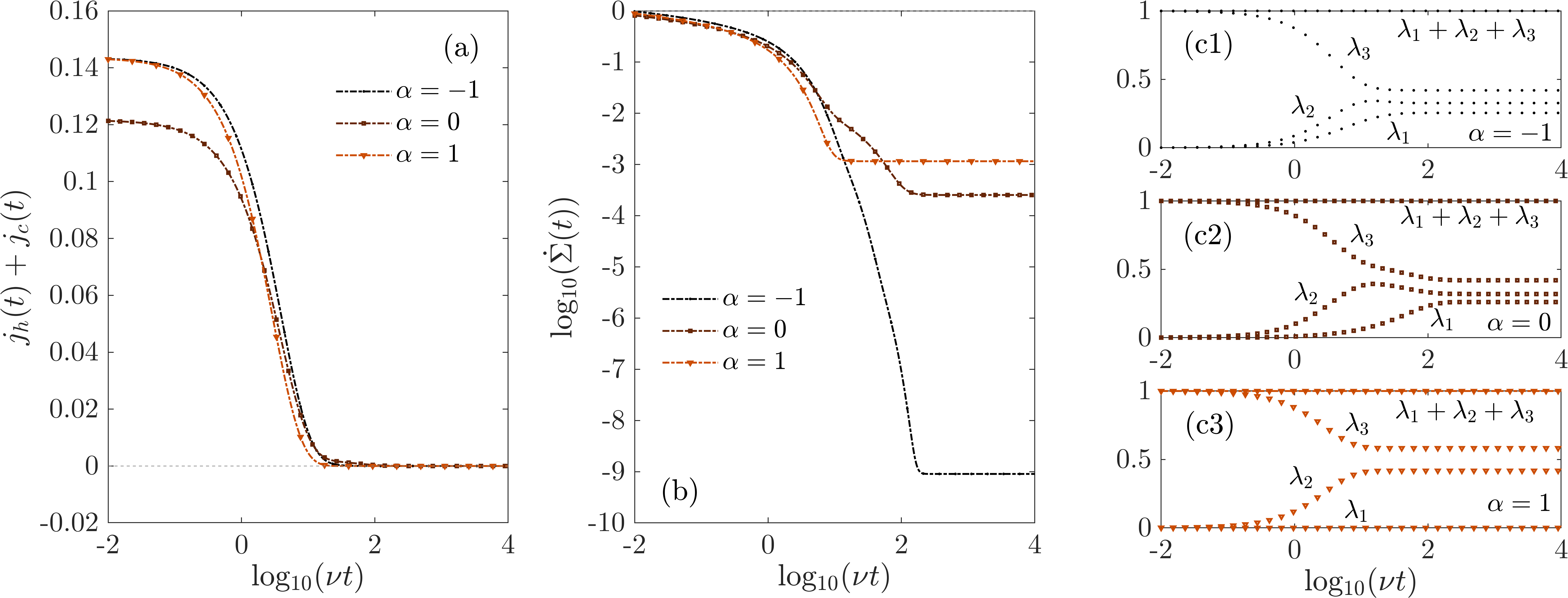} %{Figures/Newfig2.png} %%FFF Check with larger delta
\caption{Examination of thermodynamic consistencies in our work. (a) Energy conservation: The sum of heat currents $j_h(t)+j_c(t)$ approaches zero in the steady state limit. (b) Second law: The rate of entropy production $\dot{\Sigma}(t)=-j_h(t)/T_h-j_c(t)/T_c - \frac{d}{dt} \Tr{\sigma(t) \ln \sigma(t)}$ is always positive (note the log scale).
%with (b2) an inset focusing on steady state behaviour, 
(c1-3) Complete positivity: eigenvalues of the reduced density matrix $\sigma(t)$ demonstrate the preservation of trace and complete positivity for different values of $\alpha$ throughout the dynamical evolution.  Parameters are the same as in Fig. \ref{fig:Cohdyn}.}
\label{fig:fig12}
\end{figure*}
% 
%===================

\subsection{Thermal rectification}
% DDD2 unclear what is asymmetrically direction-dependent
Thermal rectification is a transport phenomenon whereby the magnitude of the heat current is different upon exchanging the direction of the applied temperature bias.
In the quantum domain, early proposals for a thermal diode (rectifier) were built on the nonequilibrium spin-boson model \cite{SpinBoson2005,Segal06}.
% heat flows is asymmetrically direction-dependent under temperature gradients, allowing applications e.g., insulative and conductive properties on opposite sides \cite{R1,R2}. 
Recently, a quantum thermal diode was experimentally realized using superconducting quantum circuits \cite{Pekola2018}.

The diode effect is realized in systems that include  (i) anharmonicity and (ii) asymmetry \cite{Pekola2018,R1,SpinBoson2005,Segal06,SB3}.
Anharmonicity is captured in the V model due the three-level structure, distinct from the harmonic oscillator spectrum of the baths.
Spatial asymmetry is introduced into the V model due to the imbalance between the arms,  courtesy of the coherence parameter $\alpha$ \cite{SpinBoson2005}. 
The figure of merit used to quantify the thermal diode effect is the rectification ratio, $R = \abs{\frac{j_h(\Delta T)}{j_h(-\Delta T)}}$, with rectification marked by $R \neq 1$.

For quasi-degenerate levels, $\Delta \to 0$, 
%NAS I think it will act as a diode for any \Delta
the nonequilibrium V system acts as a thermal diode once $\alpha\neq \pm1$:
For $\alpha=1$ (see Fig. \ref{fig:diagramp}), the model is spatially symmetric and thus trivially it cannot act as a diode. %%FFF isnt the a=-1 also spatially symmetric?
For $\alpha=-1$, the current itself vanishes when $\Delta\to 0$.
The impact of interferences on thermal rectification and heat amplification was investigated in the fully degenerate nonequilibrium V model, by setting $\Delta$ to zero \cite{VRect}.
%NAS what happens when \Delta \neq 0? What would happen for finite delta
%
%Considering intermediate values of $\alpha$,
In contrast, in Fig \ref{fig:rect} we study the diode effect in a  nondegenerate system with tunable $\alpha$. We observe the following:

First,  Fig \ref{fig:rect}(a) shows that  steady-state coherences are asymmetric in $\alpha$, as was also observed in Fig. \ref{fig:Fig5}. 
Specifically, there are striking differences in the behavior of coherences with $\Delta T=T_h-T_c$ between $\alpha = -0.5$ and $\alpha = 0.5$:
While in the former, coherences increase monotonically with $\Delta T$, this is not the case for the latter. 
Second, a larger temperature bias produces greater coherences in absolute value.
Third, there is an asymmetry in the magnitude of coherences with respect to temperature biases: a positive bias mostly produces larger coherences than a negative temperature bias. 
%First, we note that coherences and thus the heat current current grow monotonically 
%with the temperature bias.

Next, in  Fig \ref{fig:rect}(b1)-(b2) we focus on the heat current and the rectification ratio. 
For $\alpha = \pm 1$, no rectification is observed. For intermediate values of $\alpha$,  rectification behavior is observed, and is maximal when $\alpha = 0$. 
Regardless, rectification ratios achieved in this model are rather modest, reaching at most $R\sim 1.7$. %Observations here qualitatively correspond to that of Ref. \cite{VRect}.
It remains a challenge to realize a significant nanoscale quantum thermal rectifier.

%===============================================
\subsection{Thermodynamic consistency of the Redfield equation for the V model}

Quantum master equations in the Redfield form are known to exhibit pathologies \cite{RP3,RP2,RP1}. Most notably, the Redfield equation is in principle not a completely-positive and trace preserving map \cite{Strunz2021}, allowing  violations of thermodynamic laws as observed in the transient domain \cite{SL2,SL1}.
To support our analysis in this paper, in Fig. \ref{fig:fig12} we study properties of the Redfield quantum master equation for the V model, demonstrating numerically the thermodynamic consistency of our results.

First, in Fig. \ref{fig:fig12}(a) we test the conservation of energy by studying the sum of the two heat currents, $j_h(t) + j_c(t)$
 for different values of $\alpha$. 
Recall our sign convention:  currents are positive when flowing towards the system.
In the steady state limit, we find that this sum approaches zero,  validating the first law of thermodynamics of energy conservation.
Next, in Fig. \ref{fig:fig12}(b) we test the second law by displaying the entropy production rate $\dot{\Sigma}(t)$ as a function of time,
$\dot{\Sigma}(t) = -\frac{j_h(t)}{T_h} -\frac{j_c(t)}{T_c} - \frac{d}{dt} \Tr{\sigma(t) \ln\sigma(t)}$. 
Throughout the time evolution, the entropy production rate is positive, and thus the system obeys the second law. 
Finally, in Fig. \ref{fig:fig12} (c1)-(c3) we test whether for the V model the Redfield equation is a completely-positive trace-preserving map. 
We find that
throughout the dynamical evolution, the eigenvalues of the reduced density matrix are strictly positive, ensuring complete positivity. The sum of the eigenvalues is unity, implying trace preservation. Pathologies associated with the Redfield equation therefore do not manifest in the V model for the studied parameters.

%======================================
\section{Summary}
\label{S:Summary}

We studied a quantum thermal junction in which quantum coherences play a decisive role. While in the popular nonequilibrium spin-boson nanojunction coherences in the system do not alter transport %insignificant 
(in the weak system-bath coupling limit), in the  V model coherences are generated by the nonequilibrium heat baths, and
their transient and steady state behavior go hand in hand with that of the  heat current.
Our main achievements are:

(i) In the limit of quasi-degenerate excited states we used the nonsecular Redfield equation and derived closed-form expressions in steady state identifying conditions for nonzero coherences and heat currents with a tunable interference parameter $\alpha$.
We found that quantum coherences were non-vanishing in the steady-state limit due to a combination of coherent and incoherent factors: (a) the V model was in a nonequilibrium steady-state, (b) the parallel transport
pathways did not destructively interfere, and (c) the parallel pathways did not cancel out due to an effective equilibrium condition on one of the arms. % generated  when nonequilibrium conditions intermingle with the coherence parameter.
Quantum coherences were generated and sustained in our model due to the incoherent baths,
and this method may find practical applications 
in emerging quantum information technologies. % as an improvement to conventional qubits. 
% Further investigations in the strong coupling limit with techniques such as the Reaction Coordinate Quantum Master Equation method, are to be examined elsewhere.

(ii) The Redfield approach is more cumbersome to handle than the Lindblad equation, and in principle is not always granted the luxury of  complete positivity. Here we numerically showed that for the V model and the examined parameters, the Redfield map was completely positive and trace preserving.

%numerical simulations show such problems do not arise in the regime we use here. 
(iii) By studying the behavior of the heat current under the secular approximation we identified physics missed by making this common approximation. 

(iv) While our focus has been on the steady-state limit, we had further performed numerical simulations in the transient domain.
We found that the behavior of coherences in  the transient regime was markedly different compared to the steady state limit, indicating that different physical factors were at play. Indeed, while the timescale of the transient dynamics is inversely related to $\Delta^2$, see Refs \cite{Timur14,b1,b2,b3,b4,b5,comment,dodin_quantum_2016}, % DDD3
%our paper noneq eq Brumer
% as we showed for $\alpha=1$, % DDD3?
steady state coherences, population and currents only secondarily-negligibly depended on $\Delta$ in the quasi-degenerate limit.

In sum, we derived a closed-form expression for the heat current, Eq. (\ref{eq:fsscur}), which depended on the quantum interference control parameter ($\alpha$).
%thus reflecting the effect of quantum interferences on thermal transport. 
We expect that this analytical result will serve as a starting point for additional
investigations of quantum coherence effects in thermal transport.
Open experimental challenges include the construction of phase-controlled quantum thermal devices realizing this system with generalized complex couplings \cite{Giazotto}. 
From the theory side,  open questions concern
 the interplay of quantum coherences and non-Markovianity, understanding the impact of strong system-bath coupling on quantum coherences,
% (3) designing a heat engine compounding on the V model, which benefits on quantum coherences, 
and  uncovering the relationship between quantum coherences and fluctuations.

%===============================
\begin{acknowledgments}
DS acknowledges the NSERC discovery grant and the Canada Research Chair Program. 
The work of FI was funded by the University of Toronto Excellence Award.
We acknowledge discussions with Marlon Brenes.
\end{acknowledgments}

%=================================================================

\vspace{5mm}
\begin{widetext}
%=======================
% Appendix A: more detals on the Redifeld equation, can add more on the secular limit

\renewcommand{\theequation}{A\arabic{equation}}
\setcounter{equation}{0}
\setcounter{section}{0} % reset counter
\section*{Appendix A: Redfield equation for the V model} %reduced density matrix} 
\label{App:EOM}

The general form of the Redfield quantum master equation is given in 
Eq. (\ref{eq:Redfield}). For the V model as presented in Eqs. (\ref{eq:Hs})-(\ref{eq:Htot}),
the surviving terms are
\bea
\dot{\sigma}_{32}(t) &=& -i\Delta\sigma_{32}(t)  +\sum_{m}\Big[
[R^m_{12,31}(\omega_{13})+R^{*,m}_{13,21}(\omega_{12})]\sigma_{11}(t)
\nonumber\\
&&
-[R^m_{31,13}(\omega_{31})
+R^{*,m}_{21,12}(\omega_{21})]\sigma_{32}(t)
- R^m_{31,12}(\omega_{21})\sigma_{22}(t) 
-  R^{*,m}_{21,13}(\omega_{31})\sigma_{33}(t)
\Big],
\label{eq:red32}
\eea%FFFOK
\bea
\dot{\sigma}_{22}(t) &=&
\sum_{m}\Big[ [R^m_{12,21}(\omega_{12})+R^{*,m}_{12,21}(\omega_{12})]\sigma_{11}(t) 
\nonumber\\
&-& [R^m_{21,12}(\omega_{21})
+R^{*,m}_{21,12}(\omega_{21})]\sigma_{22}(t)
-R^m_{21,13}(\omega_{31})\sigma_{32}(t)
-R^{*,m}_{21,13}(\omega_{31})\sigma_{23}(t)
\Big],
\label{eq:red22}%FFFok
\eea
and
\bea
\dot{\sigma}_{33}(t)&=&
\sum_{m}\Big[[R^m_{13,31}(\omega_{13})+R^{*,m}_{13,31}(\omega_{13})]\sigma_{11}(t) 
\nonumber\\
&-& [R^m_{31,13}(\omega_{31})
+R^{*,m}_{31,13}(\omega_{31})]\sigma_{33}(t)
-R^m_{31,12}(\omega_{21})\sigma_{23}(t)
-R^{*,m}_{31,12}(\omega_{21})\sigma_{32}(t)
\label{eq:red33} \Big], %%%FFFOK
\eea
where $m=h,c$ is the index for the heat bath and the frequencies are calculated from the eigenenergies of the system, $\omega_{ij}=E_i-E_j$. Note that by construction, there are no coherence terms between the ground state and the excited states.
To calculate the dissipator, we use Eq. (\ref{eq:Rabcd}) 
ignoring the so-called Lamb shifts, an approximation that  should be valid when $T>\Delta$.
We further make use of the detailed-balance relation and write all rates in terms of bath-induced decay rates, 
$k(\omega)$ with $\omega>0$ [see Eq. (\ref{eq:rate})].
The coherences satisfy
\bea
\dot{\sigma}_{32} (t)= &-&i\Delta  \sigma_{32}(t)
-\frac{1}{2}\left[k_h(\omega_{31}) + k_h(\omega_{21}) 
+ \alpha^2 k_c(\omega_{31}) + k_c(\omega_{21})   \right] \sigma_{32}(t)
\nonumber\\
&-&\frac{1}{2} \left[ k_h(\omega_{21})  + \alpha k_c(\omega_{21}) \right] \sigma_{22}(t) 
-\frac{1}{2} \left[ k_h(\omega_{31})  + \alpha k_c(\omega_{31}) \right] \sigma_{33}(t)
\nonumber\\
&+&\frac{1}{2}
\left[
k_h(\omega_{31}) e^{-\beta_h\omega_{31}}
+k_h(\omega_{21}) e^{-\beta_h\omega_{21}}
+ \alpha k_c(\omega_{31}) e^{-\beta_c\omega_{31}}
+\alpha k_c(\omega_{21}) e^{-\beta_c\omega_{21}}
\right] \sigma_{11}(t).
\label{eq:red32A} %FFF OK
\eea
The excited state population dynamics includes gain-loss terms from-to the ground state, 
as well as the coupling to coherences,
\bea
\dot{\sigma}_{22}(t)&=&
-\left[k_h(\omega_{21}) + k_c(\omega_{21})\right] \sigma_{22}(t)
+\left[k_h(\omega_{21})e^{-\beta_h\omega_{21}}+  k_c(\omega_{21})e^{-\beta_c\omega_{21}} 
\right] \sigma_{11}(t)
\nonumber\\
&&-
\left[k_h(\omega_{31}) + \alpha k_c(\omega_{31})
\right] \sigma_{32}^R(t).
\label{eq:red22A} %FFF OK
\eea
A similar equation of motion holds for the other excited state,
\bea
\dot{\sigma}_{33}(t)&=&
-\left[k_h(\omega_{31}) + \alpha^2 k_c(\omega_{31})\right] \sigma_{33}(t)
+ \left[k_h(\omega_{31})e^{-\beta_h\omega_{31}} + \alpha^2 k_c(\omega_{31})e^{-\beta_c\omega_{31}}
\right] \sigma_{11}(t)
\nonumber\\
&&-
\left[k_h(\omega_{21}) + \alpha k_c(\omega_{21})
\right] \sigma_{32}^R(t).
\label{eq:red33A} %FFF OK
\eea
We recall that these equations rely on (i) the Born-Markov approximation and (ii) the omission
of the imaginary part of the rates (``Lamb shifts"). However, unlike the Lindblad form, the equations arise from microscopic principles and they are nonsecular. 
Simulations presented in the main text are based on solving numerically 
equations (\ref{eq:red32A})-(\ref{eq:red33A}).

To solve these equations analytically,  further simplifications are necessary.
Working in the limit of nearly-degenerate excited levels, $\Delta\to0$, we make
an additional approximation for the thermally-induced decay rates, $k_h(\omega_{31})\approx k_h(\omega_{21})$, denoted in short
 by $k_h$. A similar approximation is made for $k_c$. All rates are now evaluated at the transition energy $\nu$, see Fig. \ref{fig:Fig1}.
However, the unitary term $i\Delta\sigma_{32}(t)$ in Eq. (\ref{eq:red32A}) is kept intact.
The resulting simplified equations are presented in Eqs. (\ref{eq:s23})-(\ref{eq:s33}). In steady state, after using the population normalization condition, we recast the system in an algebraic form as
% DDD3 added minus sign %%%FFFF removed minus sign but added minus sign in front of the matrix to be consistent with next appendix
% DDD4 OK
\bea
%\dot{{\vec{v}}}=
0=\mathcal{M}\vec{v}+\vec{d}, \label{diff}
\eea
with the steady state solution $\vec{v}\equiv[\sigma_{22}\ \sigma_{33} \ \sigma_{32}^{R}]^T$. The matrix is
%
%\begin{widetext}
\bea \label{mattt}
\mathcal{M}=-
\begin{bmatrix}
(e^{-\beta_h   \nu}+1)k_h+(e^{-\beta_c   \nu}+1)k_{c} & e^{-\beta_h   \nu}k_h+e^{-\beta_c\nu}k_{c} & k_h+\alpha k_c \\
e^{-\beta_h   \nu}k_h+\alpha^2e^{-\beta_c   \nu}k_{c} & (e^{-\beta_h   \nu}+1)k_h+\alpha^2(e^{-\beta_c   \nu}+1)k_{c} & k_h+\alpha k_c\\
\alpha \frac{k_c(1+2e^{-\beta_c\nu})}{2} + \frac{k_h(1+2e^{-\beta_c\nu})}{2} &\alpha \frac{k_c(1+2e^{-\beta_c\nu})}{2} + \frac{k_h(1+2e^{-\beta_c\nu})}{2} & \xi+\Delta^2/\xi
\end{bmatrix}, %FFF OK
\eea
and the constant terms are
\bea \label{dvc}
\vec{d}=
\begin{bmatrix}
e^{-\beta_h   \nu}k_h+e^{-\beta_c\nu}k_{c}\,\,\,\,\,\, & e^{-\beta_h   \nu}k_h+\alpha^2e^{-\beta_c   \nu}k_{c}\,\,\,\,\,\, & e^{-\beta_h   \nu} k_h + \alpha e^{-\beta_c   \nu}k_{c}
\end{bmatrix}^T
\eea
The steady state solution is obtained by %setting $\dot{\vec{v}}$ to zero and 
inverting the coefficient matrix $\mathcal{M}$.
%In the steady state limit, the corresponding algebraic equations can be solved analytically, as we explain in the main text. 
%%%FFF eq A8 (3x3) only valid at steady state, for dynamics in this form the matrix would be 4x4

To obtain the analogous set of \textit{secular} equations, coherent terms are set to vanish in Eqs. (\ref{eq:red22A}) and (\ref{eq:red33A}) then we proceed as with the \textit{nonsecular} analog to obtain Eq. (\ref{eq:jhs}). For completeness, we provide here expressions for population,
\bea
\sigma_{22}^{S} = \frac{(k_c \alpha^2 + k_h)(e^{-\beta_c \nu}k_c + e^{-\beta_h \nu}k_h)}{\text{Det[}\mathcal{M}^S]}
%%%FFF OK
\eea
and
\bea
\sigma_{33}^{S} = \frac{(k_c + k_h)(e^{-\beta_c \nu}k_c\alpha^2 + e^{-\beta_h \nu}k_h)}{\text{Det[}\mathcal{M}^S]} %%%FFF OK
\eea
with 
% DDD3 pls check Eq. A12 - did I mess it? %% F checked OK
\bea
\text{Det[}\mathcal{M}^S]  &=&  
\Big[k_h(1+e^{-\beta_h \nu})+k_c(1+e^{-\beta_c \nu})\Big]
%\nonumber\\
%&\times&
\Big[k_h(1+e^{-\beta_h \nu})+\alpha^2 k_c(1+e^{-\beta_c \nu})\Big] \nonumber\\
&-&\Big(k_he^{-\beta_h \nu}+k_ce^{-\beta_c \nu}\Big)
\Big(k_he^{-\beta_h \nu}
+\alpha^2 k_ce^{-\beta_c \nu}\Big),
\eea
where recall that the $S$ superscript denotes the secular limit.
%\end{widetext}
%=============================================================
 
% Appendix B
\renewcommand{\theequation}{B\arabic{equation}}
\setcounter{equation}{0}
\setcounter{section}{0} % reset counter
\section*{Appendix B: Equilibrium single-bath V model}
\label{AppB}

Here, we investigate the V model when coupled to a single bath. In this case, the system is expected to equilibrate to a Gibbs state in the long time limit. Explicitly, we set $k_c=0$ in Eqs. (\ref{eq:s23})-(\ref{eq:s33}) and $k_h=k$ as the decay rate,
\bea
\dot{\sigma}_{32}(t)&=& -i\Delta\sigma_{32}(t)
-k  \sigma_{32}(t)
- \frac{1}{2}k \left[ \sigma_{22}(t)+  \sigma_{33}(t) \right]
+ ke^{-\beta\nu} \sigma_{11}(t),
\label{eq:s23A}%FFF OK
\\
\dot{\sigma}_{22}(t)&=&
-k  \sigma_{22}(t) 
+ ke^{-\beta\nu} \sigma_{11}(t)
- k  \sigma_{32}^R(t),
\label{eq:s22A}%FFF OK
\\
\dot{\sigma}_{33}(t)&=&
- k \sigma_{33}(t) 
+ ke^{-\beta\nu}  \sigma_{11}(t)
- k \sigma_{32}^R(t).
\label{eq:s33A}%FFF OK
\eea
These equations of motion map to a limiting configuration of Eqs. (1a) and  (1b) of Ref. \cite{dodin_quantum_2016} as follows.
First, in their notation, $r=r_1=r_2$ is the pumping rate while $\gamma=\gamma_1=\gamma_2$ is the spontaneous decay rate.
%%%FFFF added limiting configuration comment as their eqs 1a and b are more general

To match the equations (note that their physical scenario is distinct),
we identify our decay rate by $k=r+\gamma$ with $r=2J(\nu)n(\nu)$ and $\gamma=2J(\nu)$.
Since $k=2J(\nu)[n(\nu)+1]$, we conclude that $r=ke^{-\beta \nu}$.
Recall that $J(\nu)$ is the spectral density of the bath and  $n(\nu)$ is the Bose-Einstein function.
Further setting $p=1$ we get 
%and $p=1$. Recall that here $k=2J(\nu)(n_B(\nu)+1)$ and $ke^{-\beta \nu}=2J(\nu)n_B(\nu)$ so $k=r+\gamma$ and $ke^{-\beta \nu}=r$.
%%FFF Dodinpaper2016 has a typo...eq1a and eq1b doesnt match to eq2...
%
\bea
\dot{\sigma}_{22} (t)&=& 
- (r+\gamma)\sigma_{22}(t)+ r\sigma_{11}(t)- (r+\gamma) \sigma_{32}^R(t)  %\label{red22one2}%FFF OK
\nonumber\\
\dot{\sigma}_{33} (t)&= &- (r+\gamma)\sigma_{33}(t)+ r\sigma_{11}(t)- (r+\gamma) \sigma_{32}^R(t)   \nonumber\\%FFF OK
\dot{\sigma}_{32} (t)&=& -(r+\gamma+i\Delta)\sigma_{32}^R(t)+r\sigma_{11}(t)
-\frac{r+\gamma}{2}(\sigma_{22}(t)+\sigma_{33}(t)).
\label{eq:red32one2}%FFF OK
\eea
These equations are indeed identical to those  presented in Ref. \cite{dodin_quantum_2016}. % DDD3 Why wasnt this paper cited above, with the rest of Paul's papers?

We now recast equation (\ref{eq:red32one2}) in a matrix form (after applying the normalization condition $1=\sigma_{11}(t)+\sigma_{22}(t)+\sigma_{33}(t)$),
$\dot{{\vec{v}}}(t)=\mathcal{M}\vec{v}(t)+\vec{d}$
with $\vec{v}(t)\equiv[\sigma_{22}(t)\ \sigma_{33}(t) \ \sigma_{32}^{R}(t) \ \sigma_{32}^I(t)]^T$, $\vec{d}\equiv r[1 \ 1 \ 1\ 0]^T$ and the matrix
\bea \label{matrixone}
\mathcal{M}=
\begin{bmatrix}
-(2r+\gamma) & -r & -(r+\gamma) & 0\\
-r & -(2r+\gamma) &  -(r+\gamma) & 0\\
-\frac{1}{2}(3r+\gamma)&-\frac{1}{2}(3r+\gamma)&-(r+\gamma)&\Delta\\
0&0&-\Delta&-(r+\gamma)\\
\end{bmatrix}. %FFF OK
\eea
This equation is {\it distinct} from Eq. (2b) of Ref. \cite{dodin_quantum_2016}, apparently due to a typo there.
%\subsection*{Dynamical solutions}

%To solve the problem in the long time limit, 
%%%FFFF same trick for dynamics so not only for steady state. It is also not an assumption, but a symmetry that is already present in the problem. This is in contrast to Paul's assumption/approximation \sigma_22=\sigma_33 (i think PRR) which also reduces the dimensionality of the problem
We make use of existing symmetry in the system by defining $P(t)\equiv\frac{1}{2}(\sigma_{22}(t)+\sigma_{33}(t))$ such that
\bea
\dot{P} (t)=- k\sigma_{32}^R(t)-\phi P(t)+(\phi-k)/2
\label{redP}
\eea
\bea
\dot{\sigma}_{32}^R(t)=-k\sigma_{32}^R(t)-\phi P(t)+\Delta\sigma_{32}^I(t)+(\phi-k)/2
\label{red32Pr}
\eea
\bea
\dot{\sigma}_{32}^I(t)=-k\sigma_{32}^I(t)-\Delta\sigma_{32}^R(t),
\label{red32PI}
\eea
where $\phi\equiv(1+2e^{-\beta \nu})k$. Note that no further approximations were made to obtain Eqs. (\ref{redP})-(\ref{red32PI}) from Eqs. (\ref{eq:s23A})-(\ref{eq:s33A}). The unique set of steady-state solutions $\sigma_{32}^R=\sigma_{32}^I=0$ and $P=\frac{\phi-k}{2\phi}$ clearly satisfies Eq. (\ref{redP})-(\ref{red32PI}) in the steady-state limit. This solution for the steady-state populations is exactly the Gibbs state with the bath's temperature.
%===============================

%======================================
\renewcommand{\theequation}{C\arabic{equation}}
\setcounter{equation}{0}
\setcounter{section}{0} % reset counter
\section*{Appendix C: The nonequilibrium V model at $\alpha=1$}
\label{AppC}

As we show in Fig. \ref{fig:Fig5} and Eq. (\ref{eq:sscohM}), out of equilibrium baths generate and sustain coherences in the excited states in the steady state limit, but this coherence nullifies when $\alpha=1$. We now show that the equations of motion at $\alpha=1$ map to the single-bath case.

We simplify Eqs. (\ref{eq:s23})-(\ref{eq:s33}) by setting $\alpha=1$,
\bea
\dot{\sigma}_{32}(t)&=& -i\Delta\sigma_{32}(t)
- \left(k_h +  k_c\right) \sigma_{32}(t)
- \frac{1}{2}\left( k_h+  k_c\right) \left[ \sigma_{22}(t)+  \sigma_{33}(t) \right]
+ \left(  k_he^{-\beta_h\nu} +  k_c e^{-\beta_c\nu}  \right)\sigma_{11}(t)
\label{eq:s23D}
\\
\dot{\sigma}_{22}(t)&=&
-\left(k_h+ k_c\right) \sigma_{22}(t) 
+\left( k_he^{-\beta_h\nu}+ k_ce^{-\beta_c \nu}\right) \sigma_{11}(t)
-\left(  k_h + k_c\right) \sigma_{32}^R(t)
\label{eq:s22D}
\\
\dot{\sigma}_{33}(t)&=&
-\left( k_h+ k_c\right) \sigma_{33}(t) 
+\left( k_he^{-\beta_h\nu}+ k_ce^{-\beta_c \nu}\right) \sigma_{11}(t)
-\left(  k_h +k_c\right) \sigma_{32}^R(t).
% DDD2 pls keep ( ) brackets
\label{eq:s33D}
\eea
We now
define $P(t)=(\sigma_{22}(t)+\sigma_{33}(t))/2$
and retrieve, remarkably, Eq. (\ref{redP})-(\ref{red32PI}) with analogous analytical solutions yet identifying $k\equiv k_h+k_c$ and $\phi\equiv(1+2e^{-\beta_h \nu})k_h+(1+2e^{-\beta_c \nu})k_c$.
We thus showed that the dynamics of the $\alpha=1$ case, with nonequilibrium baths, can be mapped to a single bath scenario (though the current in the former is non-vanishing at $\alpha=1$).

%===================

% Appendix D
\renewcommand{\theequation}{D\arabic{equation}}
\setcounter{equation}{0}
\setcounter{section}{0} % reset counter
\section*{Appendix D: Local picture for the V model}
%Site-coupled degenerate $\leftrightarrow$ Quasi-degenerate V-model}
\label{App:localglobal}

In this Appendix we provide more details on the transformation between the V model 
in the energy basis and a particular local basis picture as discussed in Ref. \cite{MK}.
%Thinking of the V model in this local basis is most instructive, 
%although there is not a unique local basis where the V model Fig. \ref{fig:Fig1} could map to with each carrying different interpretations.

In the local basis,
$|1\rangle$ denotes the ground state  and 
$|a\rangle$ and $|b\rangle$ are exactly-degenerate upper states, which are coherently coupled with
the tunneling energy $\Delta/2$.
It can be shown that the {\it local} Hamiltonian 
\bea
\hat H_S^L = 
\begin{bmatrix}
0 & 0 & 0\\
0 & \nu-\Delta/2 & \Delta/2\\
0 & \Delta/2 & \nu-\Delta/2\\
\end{bmatrix}
\eea
is diagonalized via the transformation $\hat H_S^L=\hat U \hat H_S\hat U^{\dagger}$ with the unitary 
matrix
\bea
\hat{U} = \sqrt{\frac{1}{2}}\begin{bmatrix}
1 & 0 & 0 \\
0 & -1 & 1\\
0 & 1 & 1\\
\end{bmatrix} %%%FFF OK
\eea
The eigenvalue matrix corresponds to the V model (``global") Hamiltonian Eq. (\ref{eq:Hs}). 
The system operators that couple to the baths go through the same transformation
%$S_h^L=\hat V\hat S_h \hat V^{\dagger}$ and $S_c^L=\hat V\hat S_c \hat V^{\dagger}$
yielding the system-bath coupling operators in the local basis,
% %%FF Factor of two in the transformation, talk on monday. %%changed g ket to 1 ket
\bea
\hat{S}_h^L =  \hat U\hat S_h \hat U^{\dagger} &=& |1\rangle\langle a|+|a\rangle\langle 1|,
\nonumber\\
\hat{S}_c^L = \hat U\hat S_c \hat U^{\dagger} &=& 
\frac{1+\alpha }{2}|1\rangle\langle a|+\frac{\alpha-1 }{2}|1\rangle\langle b| + h.c.,
\eea
 While the hot bath allows only the 
 transition from the ground state to $|a\rangle$, depending on $\alpha$, the cold bath may allow transitions to both excited states.
%%%FFFF added depending on alpha above

The limit $\alpha=1$ corresponds to the ``side-coupled model" as level $|b\rangle$ does not couple to the ground state.
In contrast when $\alpha=-1$ we reach the scenario of Ref. \cite{MK} where each 
excited state couples to the ground state through a different bath. In the latter case
 the tunneling energy $\Delta$ is essential for allowing energy transfer thus the
current vanishes when $\Delta \to 0$ \cite{MK}.

\end{widetext}

%----------------------------------

%\bibliographystyle{unsrt.bst} 
%\bibliographystyle{iopart-num}
%\bibliography{bibliographySigma}
\bibliography{jcoh6}

\begin{thebibliography}{97}
\bibitem [{com()}]{comment}%
  \BibitemOpen
  \href@noop {} {}\bibinfo {note} {The full dynamics of the V model in a
  nonequilibrium setting (coupled to two baths) is not fully resolved for a
  general $\alpha$. We will address this issue in our future work.}\BibitemShut
  {Stop}%
\end{thebibliography}
%%%%%%%%%%%%%%%%%%%%%%%%%%%%%%

\end{document}